\documentclass[amsmath,amssymb,nofootinbib,prd]{revtex4}
\pdfoutput=1

\usepackage{graphicx}
\usepackage{epsfig}
\usepackage{rotate}
\usepackage{amsmath}
\usepackage{amssymb}
\usepackage{amsfonts}
\usepackage{enumerate}
\usepackage{afterpage}
\usepackage{color}
\usepackage{soul}

\begin{document}

\title{Lensing and time-delay contributions to galaxy correlations}

\author{Alvise Raccanelli$^{a,b}$, Daniele Bertacca$^{c}$, Roy Maartens$^{c,d}$,
Chris Clarkson$^{e}$, Olivier Dor\'{e}$^{a,b}$\\~}

\affiliation{
$^a$Jet Propulsion Laboratory, California Institute of Technology,
Pasadena CA 91109, US \\
$^b$California Institute of Technology,
Pasadena CA 91125, US\\
$^c$Physics Department, University of the Western
Cape, Cape Town 7535, South Africa\\
$^d$Institute of Cosmology \& Gravitation, University of
Portsmouth,
Portsmouth PO1 3FX, UK \\
$^e$Centre for Astrophysics, Cosmology \& Gravitation, and, Department of Mathematics \& Applied Mathematics, University of Cape Town, Cape Town 7701, South Africa}

\begin{abstract}
Galaxy clustering on very large scales can be probed via the 2-point correlation function in the general case of wide and deep separations, including all the lightcone and relativistic effects. 
Using our recently developed formalism, we analyze the behavior of the local and integrated contributions and how these depend on redshift range, linear and angular separations and luminosity function.
Relativistic corrections to the local part of the correlation can be non-negligible but they remain generally sub-dominant. On the other hand, the additional correlations arising from lensing convergence and time-delay effects can become very important and even dominate the observed total correlation function. We investigate different configurations formed by the observer and the pair of galaxies, and we find that the case of near-radial large-scale separations is where these effects will be the most important.

\end{abstract}

\date{\today}

\maketitle


\section{Introduction}
Forthcoming optical and H{\sc i} galaxy surveys will probe the density field of the universe on scales of the order of the Hubble scale, opening up a new front for precision cosmology. We need a commensurate theoretical effort to prepare  for future surveys such as PFS~\cite{Ellis:2012rn}, DESI \cite{Levi:2013gra}, Euclid~\cite{Amendola:2012ys}, WFIRST~\cite{Spergel:2013uha} and the SKA~\cite{ska}. 
Future surveys will reach a high precision on scales that have not been probed so far -- and we require accurate theoretical modeling of the quantities that we aim to measure and the probes that we want to construct. Large scales are naturally best suited to perform tests of dark energy or modified gravity and to constrain primordial non-Gaussianity. 

Analyses of galaxy clustering have been widely used to test parameters of cosmological models (see e.g.~\cite{Samushia:2012iq, Raccanelli:2012gt, dePutter:2012sh, Zhao:2012xw, Samushia:2011cs, Reid:2012sw, Ross:2012sx, Anderson:2012sa, Sanchez:2013uxa, Cyr-Racine:2013fsa}), using different techniques. The need to include geometrical and lightcone effects in the theoretical modeling has been long recognised (see the seminal works~\cite{Nakamura:1997pi, Matsubara:1997zj, Szalay:1997cc, Matsubara:1999du, Suto:1999id, Szapudi:2004gh}). More recently, geometrical and relativistic effects have been intensively analyzed, as it becomes clear that future surveys will need very precise modeling~\cite{Papai:2008bd, Bertacca:2012tp, Raccanelli:2013dza, Yoo:2009au, Yoo:2010ni, Bonvin:2011bg, Challinor:2011bk, Bruni:2011ta, Jeong:2011as, Maartens:2012rh, Yoo:2012se, Hall:2012wd, Lombriser:2013aj, Duniya:2013eta, Camera:2013kpa, DiDio:2013sea, Bonvin:2013ogt}. 

This paper is part of a series that investigates clustering on very large scales via the 2-point correlation function~\cite{Bertacca:2012tp, Raccanelli:2013dza, Radialxi}. Here our main focus is on the integrated contributions that arise from temporal and spatial perturbations along the line of sight.
\begin{figure}[!htbp]
\centering
\includegraphics[width=6 cm]{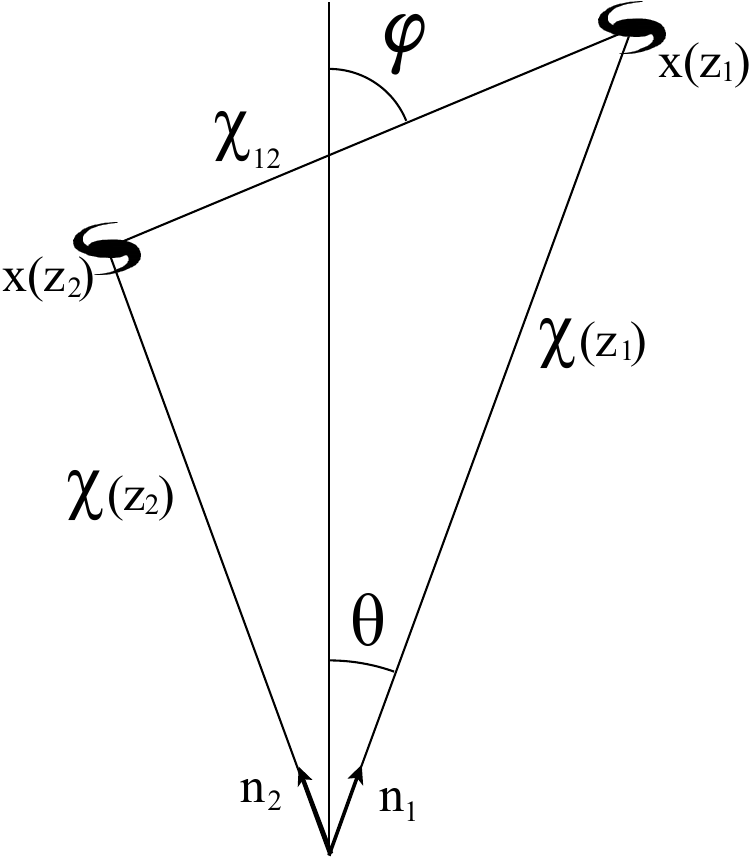}
\caption{General configuration of the system: a galaxy pair on the past light-cone, with comoving separation $\chi_{12}$, orientation angle $\varphi$ and opening angle $2\theta$.
\label{fig:triangle}
}
\end{figure}

The galaxy number density fluctuation (or the equivalent for an H{\sc i} survey) is observed on the past lightcone, and may be split into {\em local} (at the source) and  {\em integrated} (along the line of sight) parts:
\begin{eqnarray}
\label{eq:Delta}
\Delta_{\rm obs} &=& \Delta_{\rm loc}+\Delta_{\rm int} \, .
\end{eqnarray}
Schematically, we can split these parts further as:
\begin{eqnarray}
\Delta_{\rm loc} &=& b \delta + \Delta_{\rm rsd} + \Delta_{\rm wa}+\Delta_{\rm vp} \, , \\
\Delta_{\rm int} &=& \Delta_\kappa+\Delta_{\rm isw} \, .
\end{eqnarray}
In the local part, $b$ is the bias, $\delta$ is the matter overdensity, $\Delta_{\rm rsd}$ is the Kaiser flat-sky redshift-space distortion term~\cite{Kaiser:1987qv, Hamilton:1997zq}, $\Delta_{\rm wa}$ is the wide-angle contribution and $\Delta_{\rm vp}$ contains relativistic velocity and potential terms. Note that there are also relativistic contributions within $\Delta_{\rm wa}$~\cite{Bertacca:2012tp}. 

In the integrated part, the lensing magnification term is:
\begin{equation}
\label{delq}
\Delta_\kappa= 2({\cal Q}-1)\kappa \, .
\end{equation}
Here $\kappa$ is the weak lensing convergence and the magnification bias parameter ${\cal Q}$  
for a magnitude-limited survey is \cite{Jeong:2011as, Geach:2009tm}:
\begin{equation}
\label{q}
\mathcal{Q} = -{\partial \ln N_g \over \partial \ln \mathcal{L}}\bigg|_{{\cal L}={\cal L}_{\rm lim}}, 
\end{equation}
where $N_g$ is the comoving number density of galaxies of luminosity $\mathcal{L}>{\cal L}_{\rm lim}$. The precise value of $\mathcal{Q}$ depends on the survey. A detailed study of the magnification bias in future surveys is made in~\cite{Liu:2013yna}. Here we start, for simplicity, by assuming that $\mathcal{Q} = 0$, and then we look at how a range of nonzero constant ${\cal Q}$ influences the effects we consider. 
The magnification bias is often described via the logarithmic slope of the cumulative number counts, $s = \partial\ln N_g(>m)/\partial\ln m$, so that $\mathcal{Q} -1=(5 s-2) / 2$.
For intensity mapping surveys of the H{\sc I} 21cm emission, $\mathcal{Q} = 1$ and hence $\Delta_\kappa=0$~\cite{Hall:2012wd}.
 
The remaining integrated term $\Delta_{\rm isw}$  is a purely relativistic integrated Sachs-Wolfe and time-delay contribution~\cite{Yoo:2009au, Yoo:2010ni, Bonvin:2011bg, Challinor:2011bk, Jeong:2011as, Bertacca:2012tp}.

It has been recently shown that the local and integrated relativistic contributions can become significant on very large scales~\cite{Bertacca:2012tp, Raccanelli:2013dza, Yoo:2009au, Yoo:2010ni, Bonvin:2011bg, Challinor:2011bk, Bruni:2011ta, Jeong:2011as, Maartens:2012rh, Yoo:2012se, Hall:2012wd, Lombriser:2013aj, Duniya:2013eta, Camera:2013kpa, DiDio:2013sea, Bonvin:2013ogt}. Here we will investigate in detail the behaviour of the integrated terms for different cases, assuming a $\Lambda$CDM model. 
Note that these relativistic corrections apply not only in General Relativity, but also in modified gravity theories (see~\cite{Hall:2012wd, Lombriser:2013aj}).

The standard Kaiser analysis is based on $\Delta_{\rm obs}=b\delta+\Delta_{\rm rsd}$. Wide-angle corrections can be included, within a Newtonian approximation, contributing to the term $\Delta_{\rm wa}$. This was derived in~\cite{Szalay:1997cc, Matsubara:1999du} and further developed and tested in~\cite{Szapudi:2004gh, Papai:2008bd, Raccanelli:2010hk, Samushia:2011cs, Montanari:2012me, Yoo:2013zga}. The relativistic generalization of $\Delta_{\rm wa}$ was computed in~\cite{Bertacca:2012tp}.

The integrated contribution to $\Delta_{\rm obs}$ is seldom included. The effect of $\Delta_\kappa$ is typically small for the low redshifts of current surveys~\cite{Matsubara:1999du, Vallinotto:2007mf, Hui:2007cu, Hui:2007tm, Yoo:2008tj, Yoo:2009cg}. However, for very large separations we show that the lensing magnification and time-delay effects can become significant, especially in the near-radial case.

The paper is organized as follows. In Section~\ref{sec:xi} we briefly review the formalism used.
Section~\ref{sec:local_contrib} investigates the contributions of wide-angle and relativistic effects to the local part of the 2-point correlation function. 
Our main results are in Section~\ref{sec:integrated}, where we focus on the importance and behaviour of the integrated contributions to the total correlation function. In Section~\ref{sec:q} we analyze how these contributions depend on the magnification bias.
Finally, we discuss our conclusions in Section~\ref{sec:conclusions}.


\section{The 3D correlation function in the general case}
\label{sec:xi}

The observed galaxy correlation function is  (see Fig.~\ref{fig:triangle}):
\begin{equation}
\label{obsxi}
\xi( {\bf x}_1, {\bf x}_2) = \langle
\Delta_{\rm obs}( {\bf x}_1) \Delta_{\rm obs}( {\bf x}_2)  \rangle=\xi({\bf n_1},z_1; {\bf n_2},z_2)=\xi(z_1,z_2,{\bf n}_1\cdot {\bf n}_2),
\end{equation}
where ${\bf x}$ is related to the comoving distance $\chi$ by:
\begin{equation}
{\bf x}=\chi(z) {\bf n}, ~~\chi(z)=\int_0^z{d  z' \over H(z')}.
\end{equation}
We can compute \eqref{obsxi}  via the formalism of~\cite{Bertacca:2012tp}, which generalizes previous work to incorporate all relativistic effects at linear order and all geometric contributions in configurations with very large-scale separations.

We use the spherical transform of the matter overdensity in synchronous-comoving gauge to write:
\begin{eqnarray}
\label{eq:deltas}
\Delta_{\rm loc} ({\bf x}) &=& b(z) \left\{ \left[1+\frac{1}{3}\beta(z) \right]
\mathcal{A}^0_0({\bf x}) +\frac{2}{3}\beta(z) \mathcal{A}^0_2({\bf x}) + \frac{\alpha(z)\beta(z)}{\chi(z)}
\mathcal{A}^1_1({\bf x}) + \gamma(z) \mathcal{A}^2_0({\bf x}) \right\}  , \\
\label{eq:deltak}
\Delta_\kappa ({\bf x})&=& b(z) \int^{\chi}{d\tilde\chi} \; \sigma(z,\tilde{z})\left[\mathcal{A}^0_0(\tilde{\bf x})- \mathcal{A}^0_2(\tilde{\bf x})-\frac{3}{\tilde{\chi}}\mathcal{A}^1_1(\tilde{\bf x})\right] , \\
\label{eq:deltaI}
\Delta_{\rm isw} ({\bf x})&=& b(z) \int^{\chi}{d\tilde\chi} \; \mu(z,\tilde{z})\mathcal{A}^2_0(\tilde{\bf x}) ,
\end{eqnarray}
where  $\tilde\chi =\chi(\tilde z), \tilde {\bf x}={\bf x}(\tilde z)$.
The $\mathcal{A}_\ell^n$ are the spherical transforms based on Legendre polynomials  $\mathcal{P}_\ell$:
\begin{equation}
\label{eq:A-tensor}
\mathcal{A}^n_\ell ({\bf x}) = 
\int \frac{d^3k}{(2\pi)^3} (ik)^{-n}  \, \mathcal{P}_\ell ( {{\bf n}}
\cdot  {\hat{\bf k}})\exp{\left(i{\bf k} \cdot {\bf x}\right)}\;
\delta({\bf k},z) .
\end{equation}

\begin{figure}[!htbp]
\includegraphics[width=0.49 \linewidth]{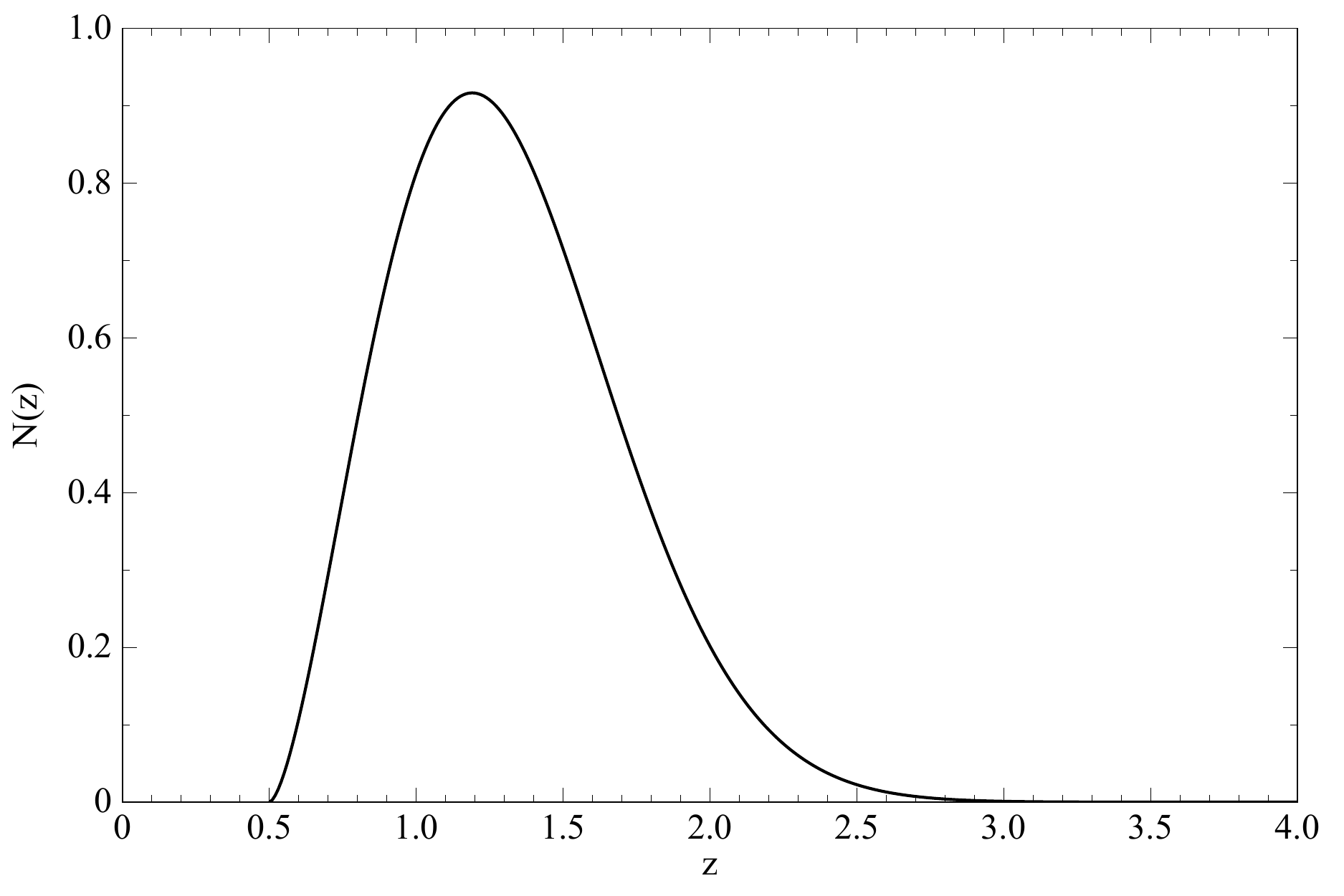}
\includegraphics[width=0.49 \linewidth]{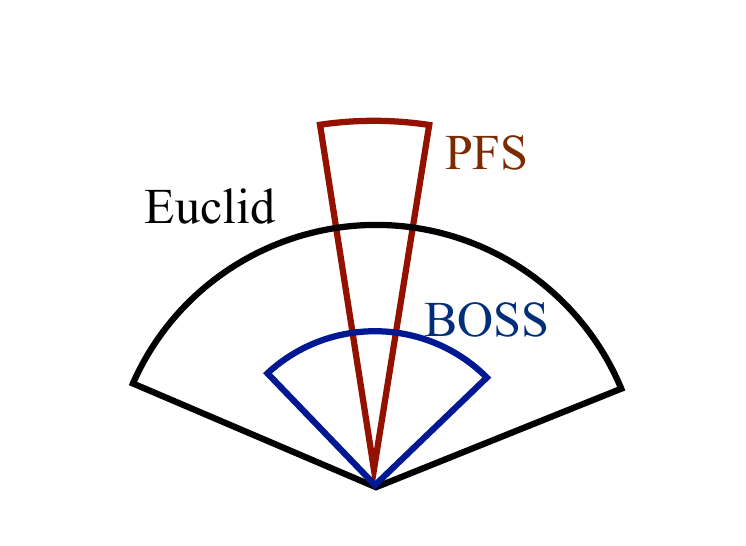}
\caption{
{\it Left:}
Normalized redshift distribution for the PFS-like survey.
{\it Right:}
Volumes probed by Euclid-like and PFS-like surveys compared to the BOSS survey.
\label{fig:nz}
}
\end{figure}

In the local overdensity, the coefficients $\alpha,\beta,\gamma$ (in the $\Lambda$CDM model) are:
\begin{eqnarray}
\label{eq:alpha}
\alpha(z)&=& - \chi(z) \frac{H(z)}{(1+z)} \left[b_e(z) -1-2\mathcal{Q}(z) +\frac{3}{2}\Omega_m (z)-
\frac{2}{\chi(z)}\big[1-\mathcal{Q}(z)\big]\frac{(1+z)}{H(z)}\right] , \\
\label{eq:beta}
\beta(z)  &=& \frac{f(z)}{b(z)}, ~~~ f:=-{d\ln D\over d\ln (1+z)}, \\
\label{eq:gamma}
\gamma(z) &=& \frac{H(z)}{(1+z)} \left\{\frac{H(z)}{(1+z)}
\left[\beta(z) -\frac{3}{2}\frac{\Omega_m (z)}{b(z)}
\right]b_e(z) + \frac{3}{2}\frac{H(z)}{(1+z)} \beta(z) \big[
\Omega_m (z) - 2 \big]
\right. \nonumber - \\
& & \left. -\frac{3}{2} \frac{H(z)}{(1+z)}  \frac{\Omega_m
(z)}{b(z)} \left[1-4\mathcal{Q}(z) + \frac{3}{2}\Omega_m
(z)\right] +\frac{3}{\chi(z)}
\big[1-\mathcal{Q}(z)\big]\frac{\Omega_m (z)}{b(z)} \right\} . 
\end{eqnarray}
The coefficients $\sigma,\mu$ in the integrated part are:
\begin{eqnarray}
\label{eq:sigma}
\sigma( z,\tilde z) &=& -2\frac{H^2(\tilde{z})}{(1+\tilde{z})^2} \frac{\left({\chi} 
-\tilde\chi\right)\tilde\chi}{ {\chi}}\frac{\big[1-\mathcal{Q}(z)\big]}{b( {z})}   \Omega_m (\tilde z) , \\
\label{eq:mu}
\mu( z,\tilde z) & = & 3
\frac{H^2(\tilde{z})}{(1+\tilde{z})^2}\frac{\Omega_m
(\tilde{z})}{b( {z})}\left\{\frac{2}{ {\chi}} \big[1-\mathcal{Q}(z)
\big] \right. \nonumber - \\
&& \left.  - \frac{H(\tilde{z})}{(1+\tilde{z})}
\big[f(\tilde{z})-1\big] \left[ b_e( {z}) - \big[1+2\mathcal{Q}(z)\big] +\frac{3}{2} \Omega_m ( {z})- \frac{2}{
{\chi}}\big[1-\mathcal{Q}( z )\big] \frac{(1+ {z})}{H(
{z})}\right]  \right\}  .
\end{eqnarray}
 Note that the simple linear bias relation $\delta_g=b\delta$ is valid near and beyond the Hubble scale only in the matter rest frame, i.e. for the comoving $\delta$   (for a careful discussion, see~\cite{Bruni:2011ta, Jeong:2011as}).
The growth rate $f$ is the quantity usually measured to constrain models of dark energy or modified gravity.  The evolution bias $b_e$ is a velocity term related to the comoving number density~\cite{Jeong:2011as}:
\begin{equation}
b_e =- {\partial \ln N_g \over \partial \ln(1+ z)}. 
\end{equation} 
For a reference survey, we take a spectroscopic survey that mimics the planned PFS (Prime Focus Spectrograph~\cite{Ellis:2012rn})  survey, which aims to measure galaxy spectra up to $z \sim 2.5$. Figure~\ref{fig:nz} shows the redshift distribution that we assume (which mimics the one predicted for PFS), as well as the cosmological volume in comparison to current surveys.
For the magnification bias \eqref{q}, we assume ${\cal Q}=0$ in the following two sections and thereafter we consider the effects of varying ${\cal Q}$.

In the standard picture of redshift-space distortions, on large scales the infall of galaxies toward higher density regions causes the well-known Kaiser effect~\cite{Kaiser:1987qv, Hamilton:1997zq}. This modifies the apparent radial position of the galaxy due to the radial component of peculiar velocities.
There is an additional effect that arises when we account for the fact that the separation angle $\theta$ is not zero ($\theta$ is neglected in the flat-sky or plane-parallel approximation). In this case the configuration loses translational invariance, giving rise to off-diagonal terms in the redshift-space distortion operator. This causes a ``mode-coupling'' effect, which is absent in the standard Kaiser analysis. For  more detail, see the following section and~\cite{Papai:2008bd, Raccanelli:2010hk}.
In a fully relativistic analysis,  further effects arise from radial (time) perturbations along the photon path.
These Doppler, Sachs-Wolfe and integrated Sachs-Wolfe effects lead to a frequency shift, so that the transformation between real- and redshift-space is modified relative to the Newtonian treatment.

\section{Local part of the correlation function}
\label{sec:local_contrib}

The local part of \eqref{obsxi} is:
\begin{equation}
\label{locc}
\xi_{\rm loc}( {\bf x}_1, {\bf x}_2)= \langle \Delta_{\rm loc}( {\bf x}_1) \Delta_{\rm loc}( {\bf x}_2)  \rangle.
\end{equation}
In this section we ignore the integrated part of $\xi$ and focus on the local part. 

We expand the correlation function in tripolar spherical harmonics $S_{\ell_1\ell_2L}$~\cite{Szalay:1997cc, Szapudi:2004gh, Papai:2008bd, Raccanelli:2010hk, Bertacca:2012tp}:
\begin{equation}\label{trip}
{\xi}_{\rm loc}( {\bf n}_1, z_1, {\bf n}_2, z_2) = b(z_1) b(z_2)
\sum_{\ell_1,\ell_2,L,n} 
B_{ n}^{\, \ell_1\ell_2L}( {\chi}_1, {\chi}_2)\, 
S_{\ell_1\ell_2L}( {{{\bf n}}}_1, { {{\bf n}}}_2,  { {{\bf n}}}_{12}) \,
\xi_L^{\, n}(\chi_{12}; z_1, z_2) ,
\end{equation}
where ${\bf x}_1- {\bf x}_2 \equiv \chi_{12} {\bf n}_{12}$. The correlation moments in this expansion are:
\begin{equation}
\label{eq:xiLn}
\xi_L^{\, n}(\chi; z_1, z_2) = \int dk \frac{k^{2-n}}{2\pi^2}
\, j_L(\chi k) \, P_\delta(k; z_1, z_2) ,
\end{equation}
where $P_\delta(k; z_1, z_2)\propto \langle \delta(k,z_1)\delta (k,z_2)\rangle$  is the matter power spectrum in synchronous-comoving gauge and $j_L$ are spherical Bessel functions.
The $B$ coefficients in \eqref{trip} contain the functions $ \alpha, \beta, \gamma$ of \eqref{eq:alpha}--\eqref{eq:gamma} (the full expressions are given in~\cite{Bertacca:2012tp}).
The correlation function in this case depend on three variables that can be chosen as any combination of angles and lengths in Figure~\ref{fig:triangle}. Here we use statistical isotropy to write $\xi(z_1,z_2,\theta)$ and we fix 2 of the 3 degrees of freedom.

The meaning of the functions $ \alpha, \beta, \gamma$  is as follows:
\begin{itemize}
\item
 $\beta$ encodes the effect of the matter overdensity $\delta$ and redshift-space distortions. The first two terms ($\beta$ terms) in~\eqref{eq:deltas} appear in the Kaiser flat-sky approximation.
(This Newtonian term is typically taken as ``the standard contribution".)
\item
 $\alpha$ encodes the ``mode-coupling'' effect, which mixes different modes of the wide-angle correlation. It depends on the velocity dispersion and is related to the fact that local overdensity around each galaxy can affect the ``apparent movement'' of galaxies when going from real- to redshift-space, and can be thought of as a velocity term~\cite{Szalay:1997cc, Papai:2008bd, Raccanelli:2010hk}.
In the case of very large separations, it acquires relativistic corrections~\cite{Bertacca:2012tp}. The third term ($\alpha\beta$ term) in~\eqref{eq:deltas} thus describes geometry and also relativistic corrections [see \eqref{alpha}].
The standard flat-sky (Kaiser) analysis includes only the first two terms of~\eqref{eq:deltas}. In the Newtonian wide-angle generalization of the Kaiser case, the third term of~\eqref{eq:deltas} appears, with the Newtonian approximation ${\alpha_{\rm nwt}}$ of $\alpha$~\cite{Bertacca:2012tp}: 
\begin{eqnarray}
\label{eq:alpha_nwt}
{\alpha_{\rm nwt}(z) \over  \chi(z)}&=& - \frac{H(z)}{(1+z)}\left\{b_e(z)-\frac{2}{\chi(z)}\big[1-\mathcal{Q}(z)\big]\frac{(1+z)}{H(z)}\right\}=  \frac{d \ln{N_g}}{d \chi}+ \big[1-\mathcal{Q}(z)\big]\frac{2}{\chi},\\
\label{alpha}
\alpha(z)&=& \alpha_{\rm nwt} (z) -  \frac{\chi(z) H(z)}{(1+z)} \left[\frac{3}{2}\Omega_m (z) -
1-2\mathcal{Q}(z)\right] .
\end{eqnarray}
For a different and equivalent definition of $\alpha$, see~\cite{Yoo:2012se,Yoo:2013zga}.
\item
 $\gamma$ encodes the effect of the gravitational potentials.  The last term ($\gamma$ term) in~\eqref{eq:deltas} is a purely relativistic term that is not present in the Newtonian treatment.
\end{itemize}

In order to understand which terms are more important and for which configuration, we analyze separate parts of the local correlation, where we divide $\xi_{\rm loc}$ as follows:
\begin{equation}
\label{locp}
\xi_{\rm loc} = \xi^{\beta}+\xi^{\beta\gamma}+\xi^{\alpha\beta}+\xi^{\alpha\beta\gamma}+\xi^{\gamma},
\end{equation}
where:
\begin{quote}

$\xi^{\beta}$ is the standard Kaiser flat-sky contribution, i.e. all the terms in \eqref{trip} with $B$ coefficients that include only $\beta$; 

$\xi^{\beta\gamma}$ includes all terms with coefficients that depend on both $\beta$ and $\gamma$ -- and hence it gives relativistic corrections to the Kaiser contribution~\cite{Bertacca:2012tp};

$\xi^{\alpha\beta}$ has all the terms that depend on both $\alpha$ and $\beta$ -- it therefore has the wide-angle and mode-coupling contributions, which include relativistic corrections via \eqref{alpha}~\cite{Bertacca:2012tp};

$\xi^{\alpha\beta\gamma}$ includes all terms that receive contributions from all of $\alpha, \beta$ and $\gamma$  -- and thus it gives further relativistic corrections to wide-angle and mode-coupling effects;

$\xi^{\gamma}$ is a purely relativistic contribution, which includes only $\gamma$.
\end{quote}

The relative importance of these components depend on the particular configuration, i.e. on $\{z_1, z_2, \theta\}$. The overall contribution of each part will be a weighted average of the contributions for all configurations measured in the survey. Here we study their dependence on separation angle, scale, and redshifts of the two galaxies.

\begin{figure*}[!htbp]
\includegraphics[width=0.49 \linewidth]{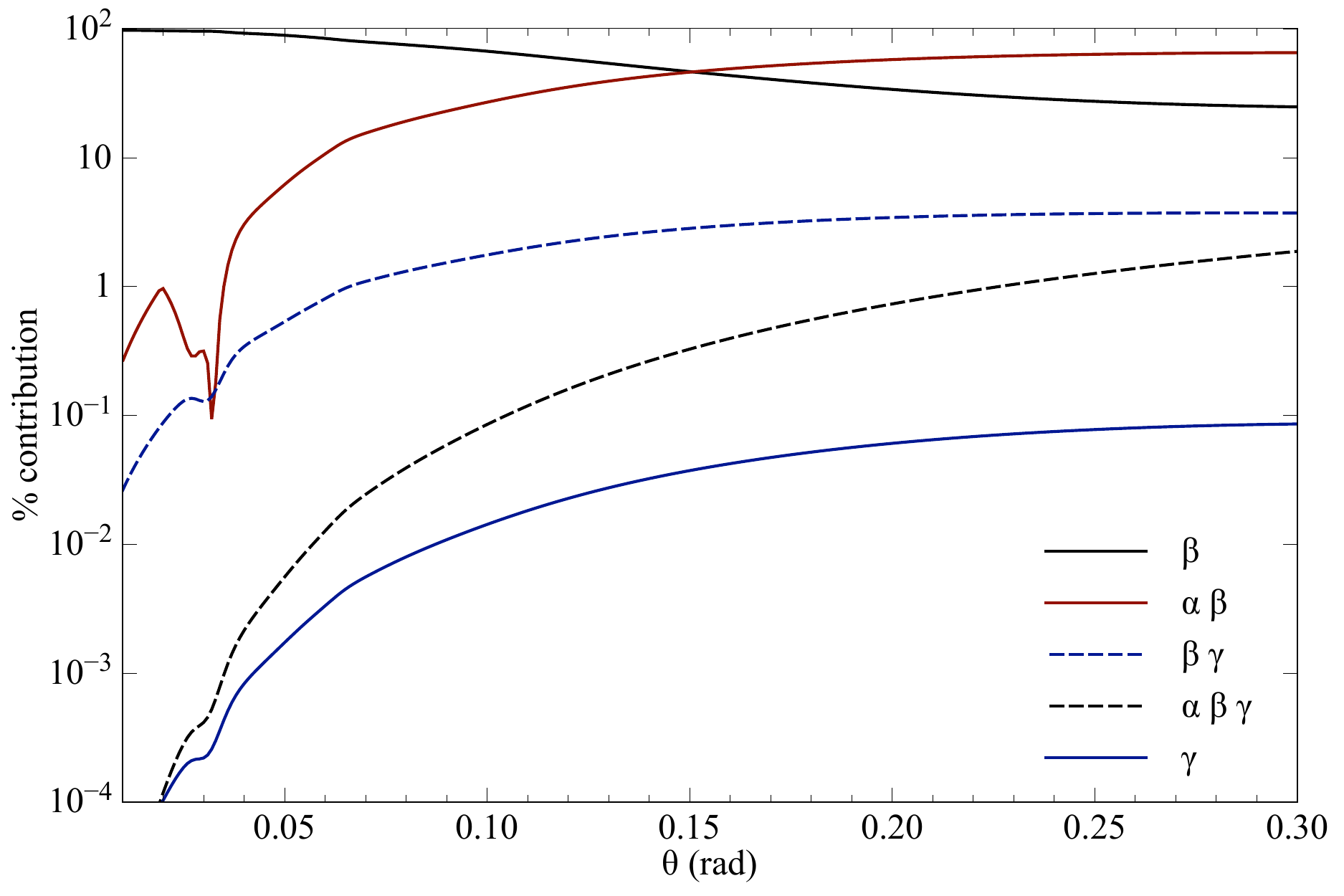}
\includegraphics[width=0.49 \linewidth]{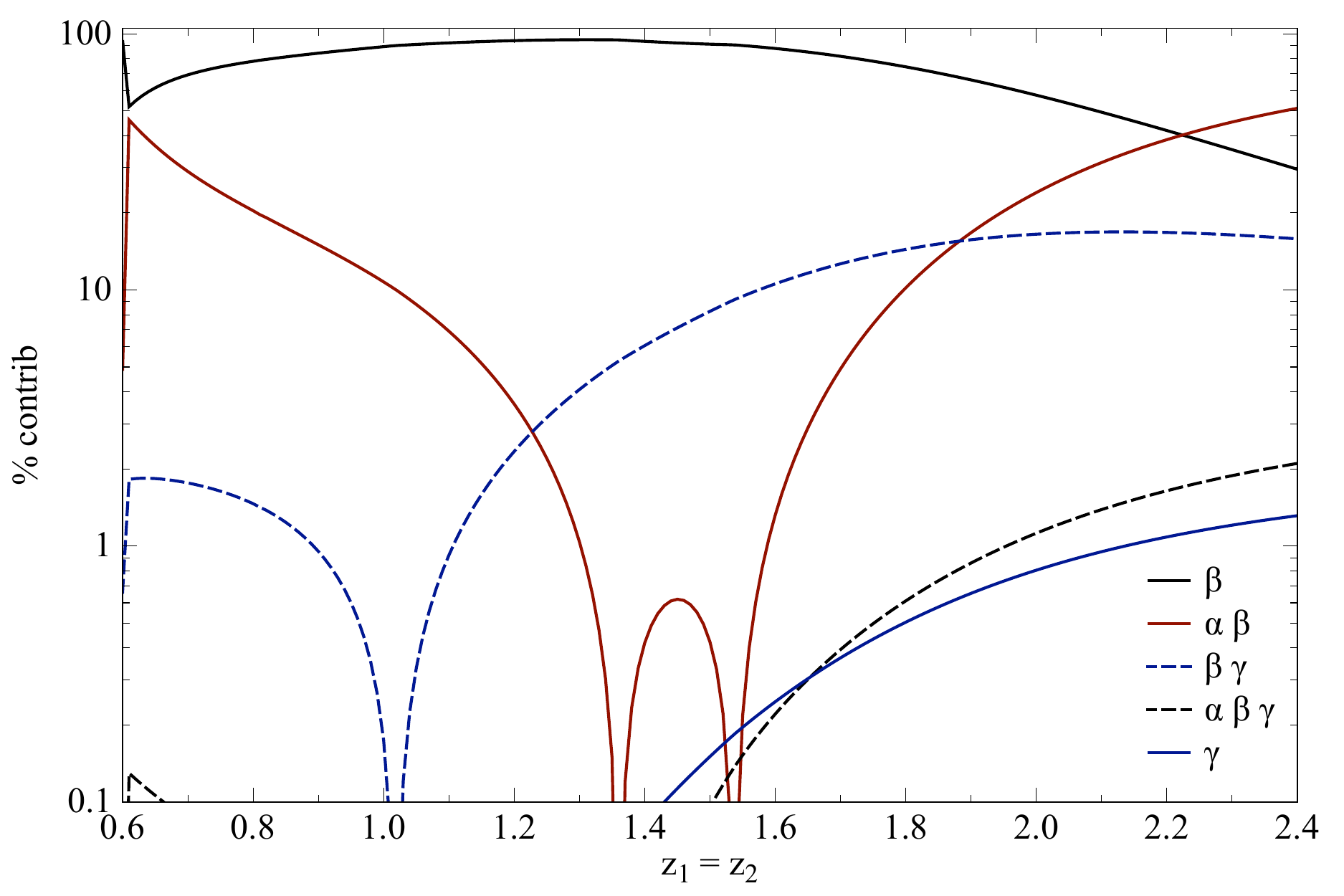}\\
\includegraphics[width=0.5 \linewidth]{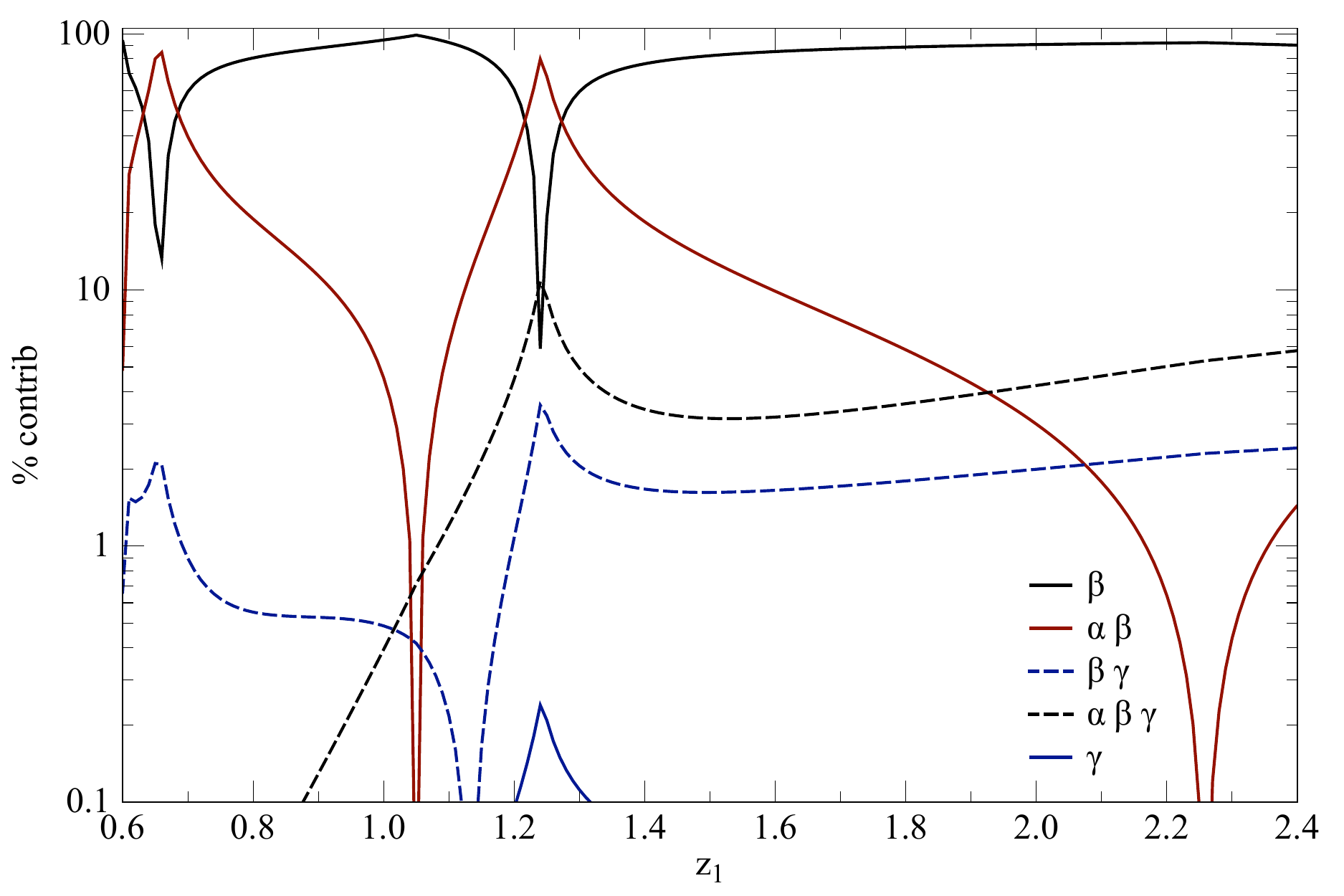}
\caption{Relative contributions to the local correlation, as in \eqref{locp}. {\em Top left:} $z_1 = z_2 = 0.7$, with $\theta$ varying. {\em Top right:} $\theta=0.1\,$rad with $z_1=z_2$ varying. {\em Bottom:} $\theta=0.1\,$rad and $z_2=0.6$ with $z_1$ varying (spikes correspond to zero-crossings).
\label{fig:theta}
}
\end{figure*}

First we consider pairs of galaxies transverse to the line of sight. Figure~\ref{fig:theta} (top left) shows how the contributions in \eqref{locp} depend on the separation angle $2\theta$. The standard Kaiser component is dominant until $\theta=0.15\,$rad and the wide-angle effects dominate at larger angular separations, as expected. The relativistic contributions do not exceed 10\% of the total local correlation $\xi_{\rm loc}$.
Figure~\ref{fig:theta} (top right) illustrates the relative contributions in the transverse case with a fixed separation angle of $2\theta = 0.2\,$rad and increasing redshift. In this case the Kaiser terms are dominant until very high redshift, with relativistic effects accounting for more than 10\%  when $z \gtrsim 1.5$.

In the case of galaxies with non-transverse separation, we fix the separation angle and one of the galaxy redshifts, and allow the redshift of the other galaxy to increase. This is shown in Figure~\ref{fig:theta} (bottom) with separation angle $\theta = 0.1\,$rad and  $z_2=0.6$. Relativistic effects become again important at high-z, in particular when $z_1 \gtrsim 1.2$.
In all cases considered, the purely relativistic contribution $\xi^\gamma$ remains negligible, but there are significant relativistic corrections to the density and wide-angle terms. 

In general, the dominant contribution to $\xi_{\rm loc}$ is from the standard Kaiser and the ``mode-coupling" parts, i.e. $\xi^\beta+\xi^{\alpha \beta}$. However, relativistic contributions to the local correlation function are non-negligible for very large separations, and will need to be taken into account if one wants a precise modeling of the local correlation function on large-scales.
The evaluation of how this impacts cosmological measurements is left to a future work.


\section{Integrated contribution to the total correlation function}
\label{sec:integrated}

The integrated contributions to the total correlation function \eqref{obsxi} are the lensing magnification term $\Delta_{\rm \kappa}$  and time-delay term $\Delta_{\rm isw}$, given by \eqref{eq:deltak}--\eqref{eq:deltaI}. They account respectively for spatial and temporal perturbations along the line of sight, modifying the apparent angular and radial positions of the source. As integrals along the line of sight, they account for the cumulative effect of the variation of gravitational potentials, so they increase with the distance between us and the source.
(It is worth emphasizing here that the $\kappa$ term is not the direct lensing of a background source by foregrounds, but is due to perturbations along the line of sight.)
The magnification bias contributes $2{\cal Q}\kappa$ towards $\Delta_\kappa$ [see \eqref{delq}], so it also modifies the apparent flux of the source. It also contributes to the time-delay term (and to some of the local redshift-space distortion effects). In Section~\ref{sec:q} we will look at how the lensing magnification and time-delay change as ${\cal Q}$ changes. Here we set ${\cal Q}=0$.
\begin{figure}[!htbp]
\includegraphics[width=0.49 \linewidth]{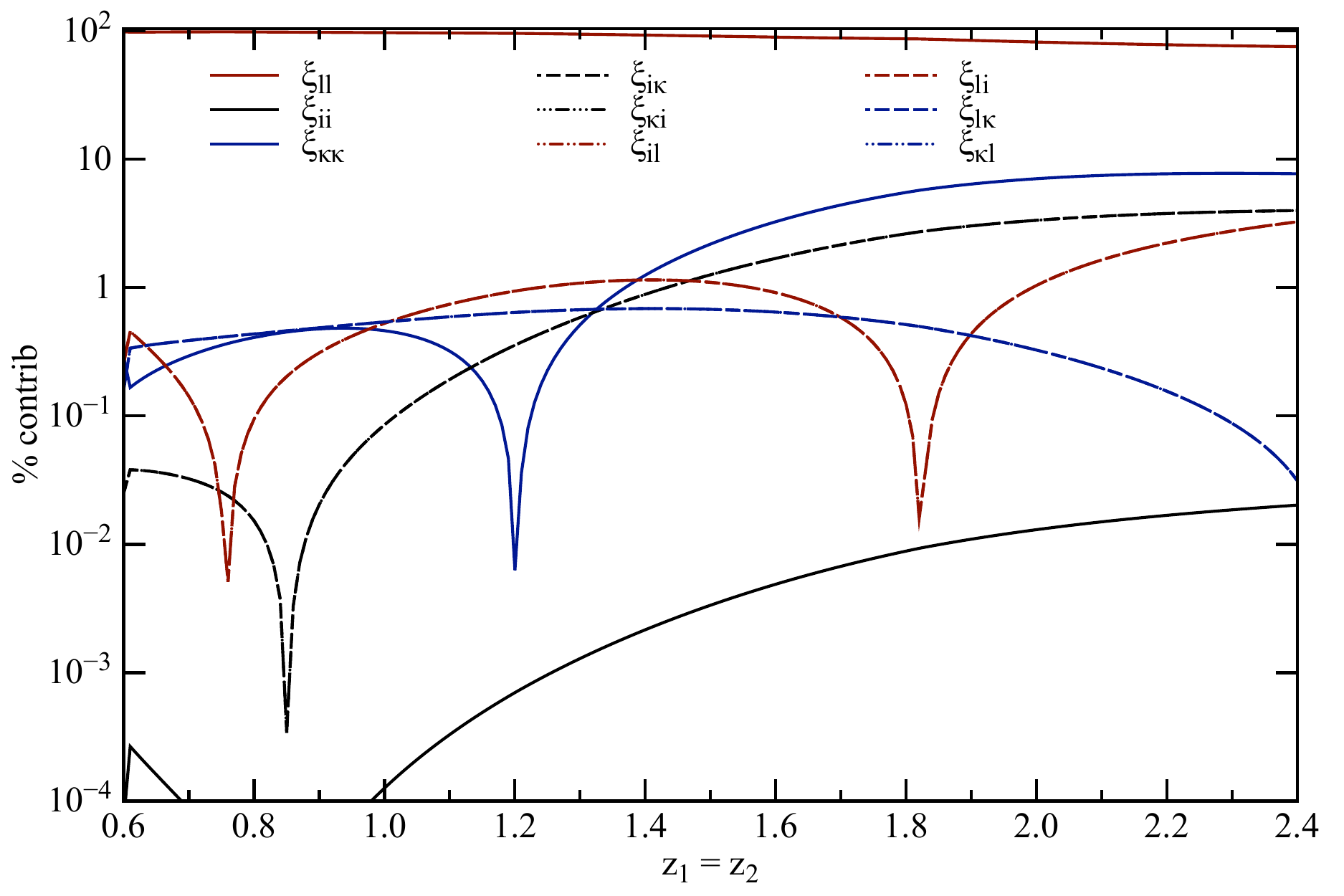}
\includegraphics[width=0.49 \linewidth]{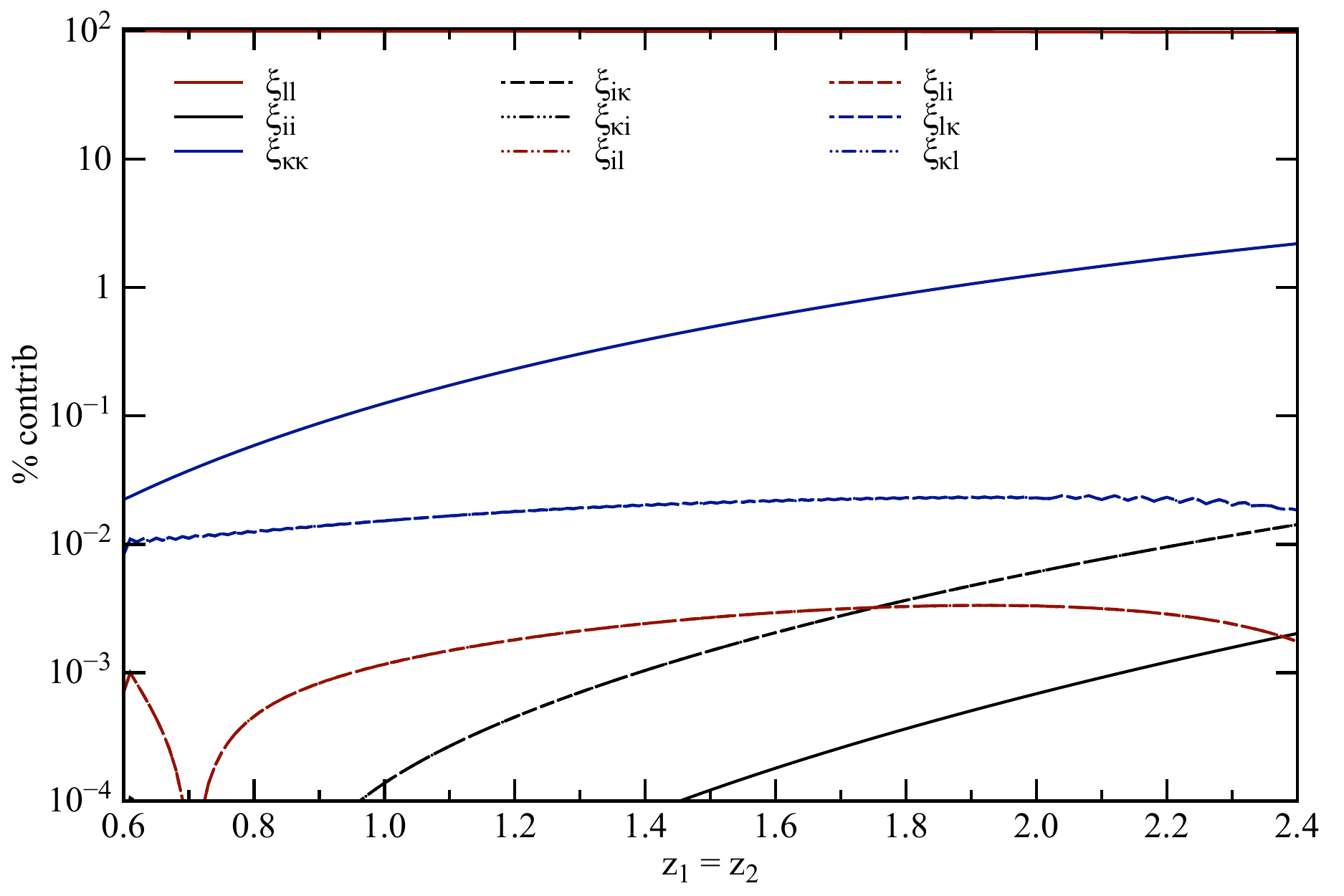}
\caption{Relative contributions of $\xi_{AB}$ to $\xi_{\rm tot}$ [see \eqref{xiab}] for transverse configurations with $z_1=z_2$,  and fixed separation half-angle of $\theta=0.1\,$rad ({\em left}) and $\theta=0.01\,$rad ({\em right}). 
\label{fig:tot_contrib_perp_t01}
}
\end{figure}
\begin{figure}[!htbp]
\includegraphics[width=0.49 \linewidth]{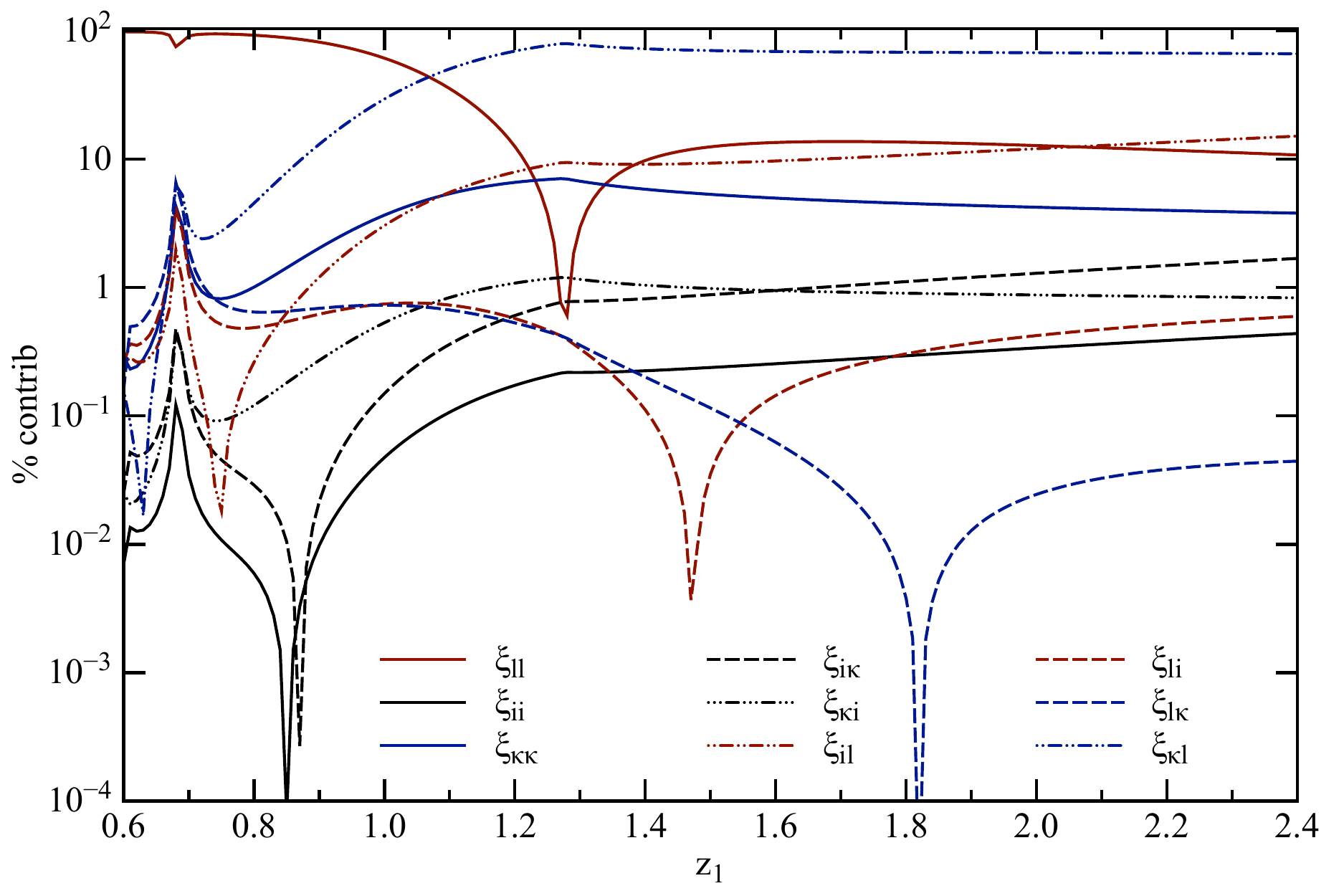}
\includegraphics[width=0.49 \linewidth]{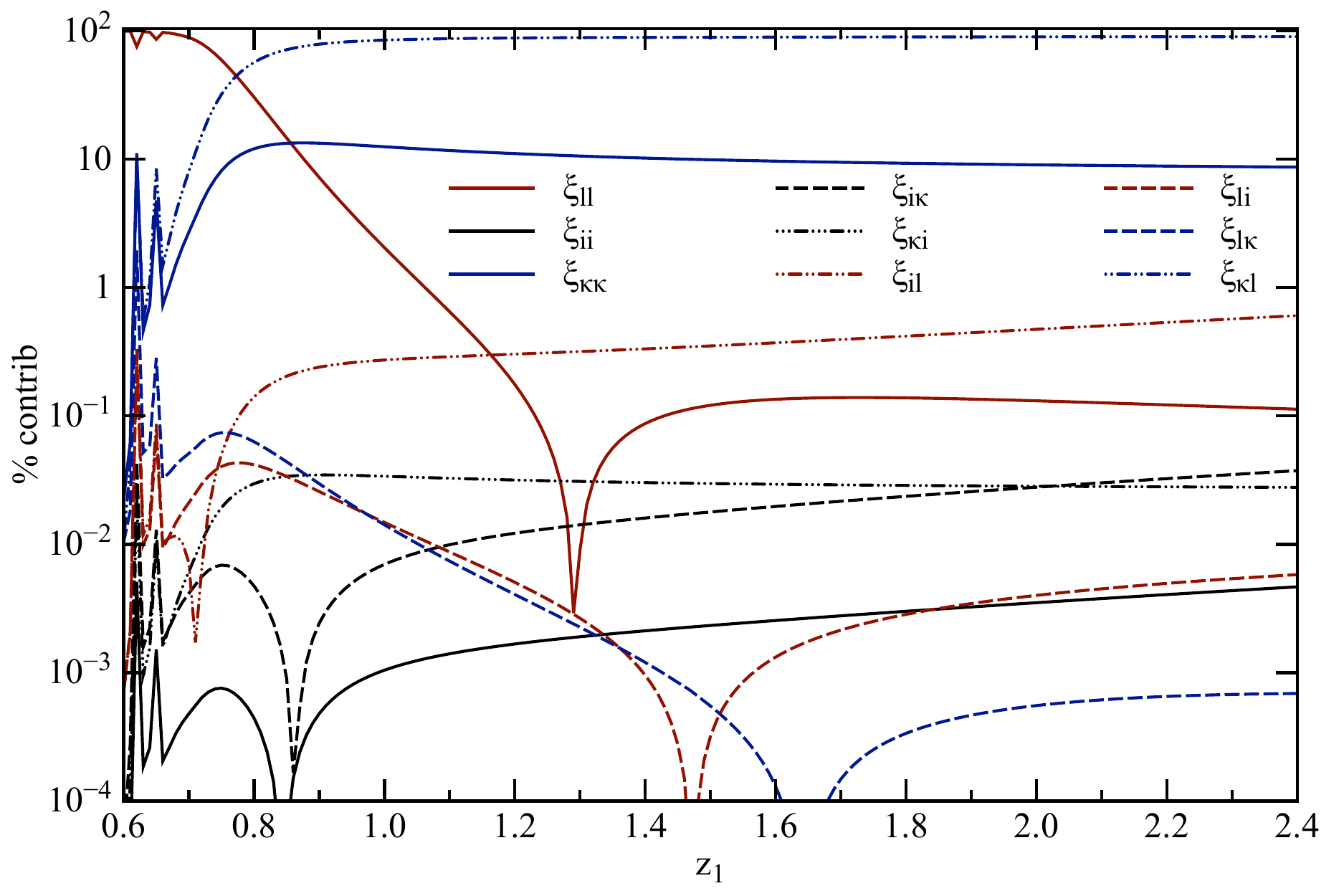}
\caption{Relative contributions of $\xi_{AB}$ for non-transverse configurations, with fixed $z_2=0.6$ and varying $z_1$. {\em Left:} with $\theta=0.1\,$rad; {\em right:}
with $\theta=0.01\,$rad.
\label{fig:tot_contrib_rad_01}
}
\end{figure}

The total {\it observed} correlation \eqref{obsxi} is a sum of correlations:
\begin{equation}
\label{xiab}
\xi_{\rm tot}=\xi_{\rm loc}+\xi_{\rm int} = \sum_{A,B} \xi_{ AB}, ~ ~\xi_{AB}=\langle \Delta_A\Delta_B\rangle,~~ A,B={\rm l}, \kappa, {\rm i} ~~~{\rm (l\equiv loc,\,i \equiv isw)}.
\end{equation}
For convenience, we have shortened loc to l and isw to i. 
The measured correlation function will be a weighted average of pairs in the $(z, \theta, \varphi )$ bin considered (see Fig.~\ref{fig:triangle}), and this is strongly survey dependent. Here we analyze different cases and probe how the relative importance of single contributions $\xi_{AB}$ depend on the configuration. 
It is worth noting that in current analyses only the first term of~\eqref{xiab}, i.e. $\xi_{\rm loc}=\xi_{\rm ll}$, is considered, while all the integrated terms, i.e $\xi_{AB}$ where at least one of $A,B$ is $\kappa$ or i, are neglected. This does not introduce considerable errors when measuring correlations on scales probed so far, but as we will show, the integrated terms can substantially contribute to the total correlation when probing larger cosmological volumes.

Figures~\ref{fig:tot_contrib_perp_t01} and \ref{fig:tot_contrib_rad_01} show the percentage contribution of all the $\xi_{AB}$ terms to the total observed correlation function~\eqref{xiab}. In the plots, the first and second subscripts refer to the galaxies at $z_1$ and $z_2$, respectively, since $\xi_{AB}(z_1,z_2,{\bf n}_1\cdot{\bf n}_2)=\langle \Delta_A(z_1,{\bf n}_1) \Delta_B(z_2,{\bf n}_2)\rangle$.

\begin{figure}[!htbp]
\includegraphics[width=0.49 \linewidth]{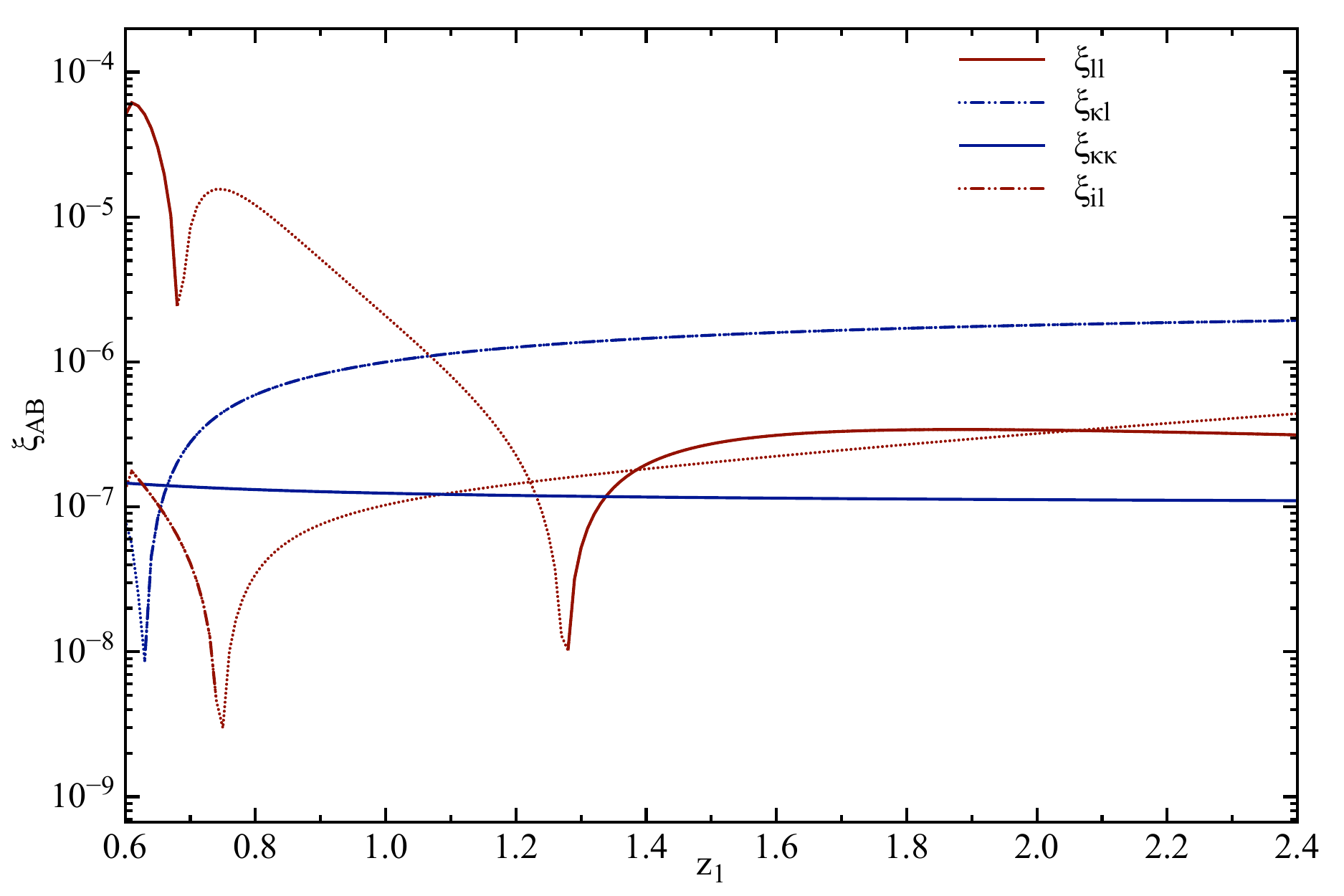}
\includegraphics[width=0.49 \linewidth]{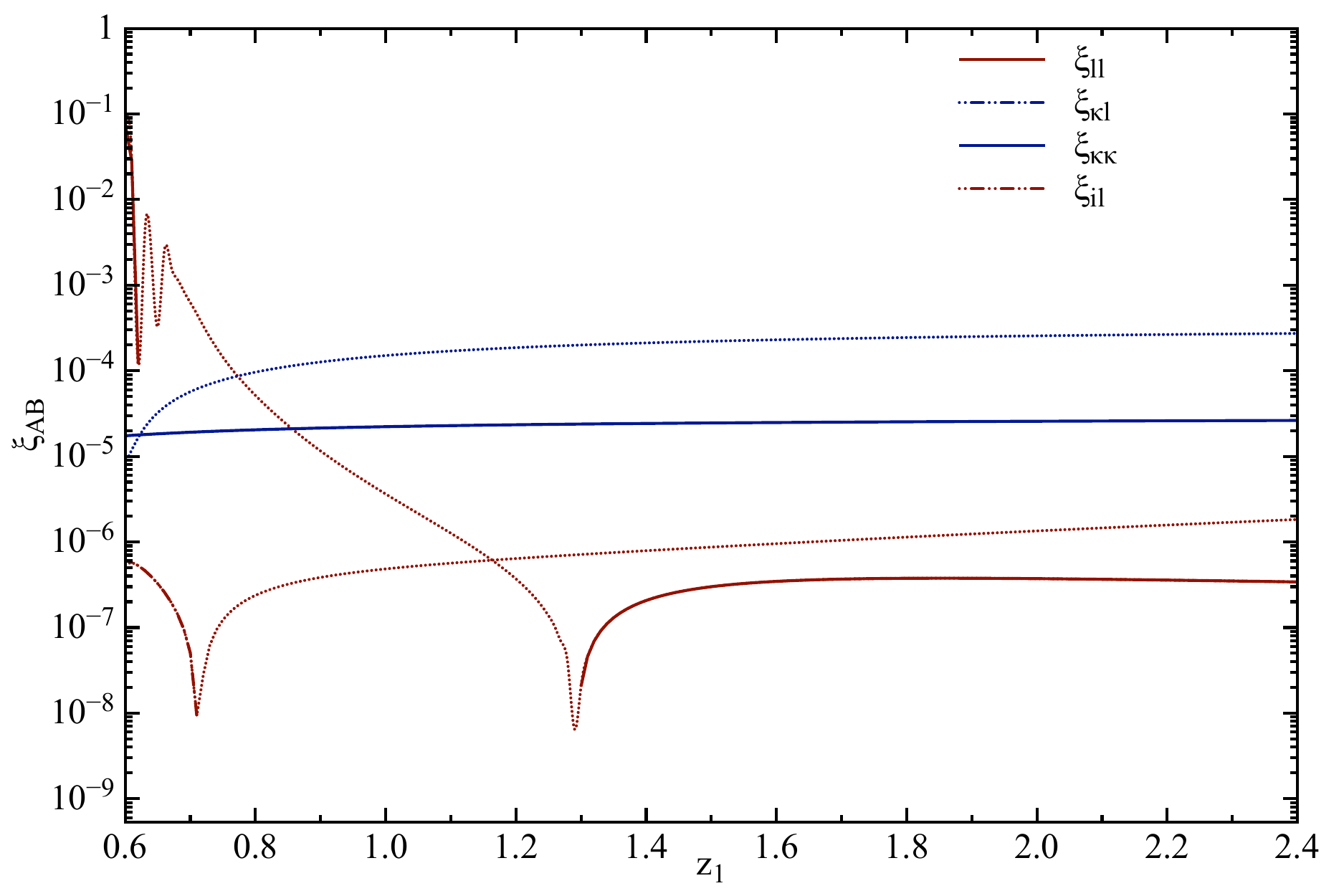}
\caption{
Absolute value of dominant terms in the observed correlation function in the non-transverse configuration; dotted parts of the lines correspond to negative vales of the correlations $\xi_{AB}$. {\em Left:} with $\theta=0.1\,$rad; {\em right:}
with $\theta=0.01\,$rad.
\label{fig:abs_xi}
}
\end{figure}

In Figure~\ref{fig:abs_xi} we plot the absolute value of the dominant terms of the total observed correlation, in order to better visualize the relative importance of the different contributions. In the plot, dotted lines represent negative correlations.

We investigate two types of configurations:
\begin{quote}
{\em Transverse} separations, with pairs across the line of sight. We fix $\theta$ and vary $z_1=z_2$ (Fig.~\ref{fig:tot_contrib_perp_t01}).\\
{\em Non-transverse} separations, including cases that are near-radial (nearly along the line of sight). We fix $\theta, z_2$ and vary $z_1$ (Figure~\ref{fig:tot_contrib_rad_01}).
\end{quote}

The key points arising from the plots are as follows.\\

In the {\it transverse} case:

\begin{itemize}
\item Figure~\ref{fig:tot_contrib_perp_t01} (left) shows that with a relatively large angular separation, the local correlation $\xi_{\rm ll}$ dominates the signal strongly up to moderately high redshifts $z\lesssim 1.4$. The integrated contribution contributes a non-negligible $\sim 1-10\%$ for $z \gtrsim 1.4$, and is dominated by $\xi_{\kappa\kappa}$, $\xi_{\rm i\kappa}$, $\xi_{\rm il}$. Note that in this case $\xi_{AB} = \xi_{BA}$.

\item
$\xi_{\rm i\kappa}$, $\xi_{\rm il}$ are purely relativistic terms -- i.e., they would not be present in a Newtonian analysis. There are also relativistic corrections to any terms including a correlation with $\Delta_{\rm l}$, including to $\xi_{\rm ll}$ (as discussed in the previous section). The other integrated term $\xi_{\kappa\kappa}$ is not purely relativistic -- but it is not included in the standard Newtonian analysis.

\item
With a small separation angle, the galaxies are relatively close together and so the local correlation is large, so that the integrated contribution is a very small fraction of the total (right panel). Conversely, for large separation angle, the local correlation is weak so that the integrated contribution is a larger fraction of the total (left panel).  The largest integrated contribution$\xi_{\kappa\kappa}$ in the right panel only reaches $\sim 1\%$ at $z \gtrsim 1.8$. The purely relativistic terms $\xi_{\rm iA}$,$\xi_{\rm Ai}$ are negligible.

\end{itemize}

In the {\it non-transverse} case:

\begin{itemize}

\item
Figure~\ref{fig:tot_contrib_rad_01} (left) shows that with the same large angular separation, but a growing redshift separation, the integrated effects become relevant even at smaller redshifts. Indeed, $\xi_{\rm \kappa l}$ and $\xi_{\rm il}$ (purely relativistic term) dominate over the local contribution for $z_1 \gtrsim 1.2$. Here the mixed lensing-local and time delay-local terms become dominant as one of the galaxies in the pair gets to high redshift. The lensing-lensing correlation becomes relevant but does not contribute more than $10\%$.

\item
With a small separation angle, i.e. near-radial separations,
Fig.~\ref{fig:tot_contrib_rad_01} (right) shows that the lensing contributions $\xi_{\rm \kappa l}$ and $\xi_{\kappa\kappa}$ quickly (for $z\gtrsim 0.8$) dominate the local-local contribution $\xi_{\rm ll}$, which decays and becomes negligible when the correlation tends to radial.

\item
The asymmetry in the non-transverse case between the correlations $\xi_{\kappa {\rm l}}$, which becomes dominant, and $\xi_{{\rm l}\kappa}$, which remains negligible, arises by the choice of which galaxy redshift to fix. We have:
\begin{equation}
\xi_{\kappa {\rm l}}(z_1,z_2)=\langle\Delta_\kappa(z_1)\Delta_{\rm l}(z_2)\rangle,
\end{equation}
where we suppressed the angular dependence for convenience. We fixed the galaxy at $z_2$, while the galaxy at $z_1$ moves further away. So we expect a strong signal from correlating the local contribution $\Delta_{\rm l}(z_2)$ and the growing lensing contribution $\Delta_\kappa(z_1)$. By contrast, $\xi_{{\rm l}\kappa}$ will be suppressed since the local contribution weakens as $z_1$ grows and the lensing contribution at $z_2$ is small and unchanging.

\item
There are contributions from all the terms in the general non-transverse case, but it is striking that near-radial correlations are enhanced and dominated by the integrated terms. The dominant contribution here is the magnification effect, as expected for long radial correlations.

\item
The radial local correlation exhibits a double dip, similar to the behavior found in~\cite{Tian:2011}. However, the two cases cannot be directly compared, as here we include relativistic terms. It is also interesting to note that the integral condition $\int_0^{\infty} \xi(r) r^2 dr = 0$ does not apply anymore, because of effects from observing on the light cone.
In fact, $\xi$ goes negative beyond the BAO peak, reaches a minimum, and then crosses zero again to become positive. This last positive correlation is due to lensing.

\end{itemize}

It has been shown (see e.g.~\cite{DiDio:2013sea}) that relativistic and lensing terms can be detected in future galaxy surveys. However, here we are setting a formalism for the full expression of the correlation function, and investigating the contribution coming from separate terms.

\begin{figure*}[!htbp]
\includegraphics[width=0.49 \linewidth]{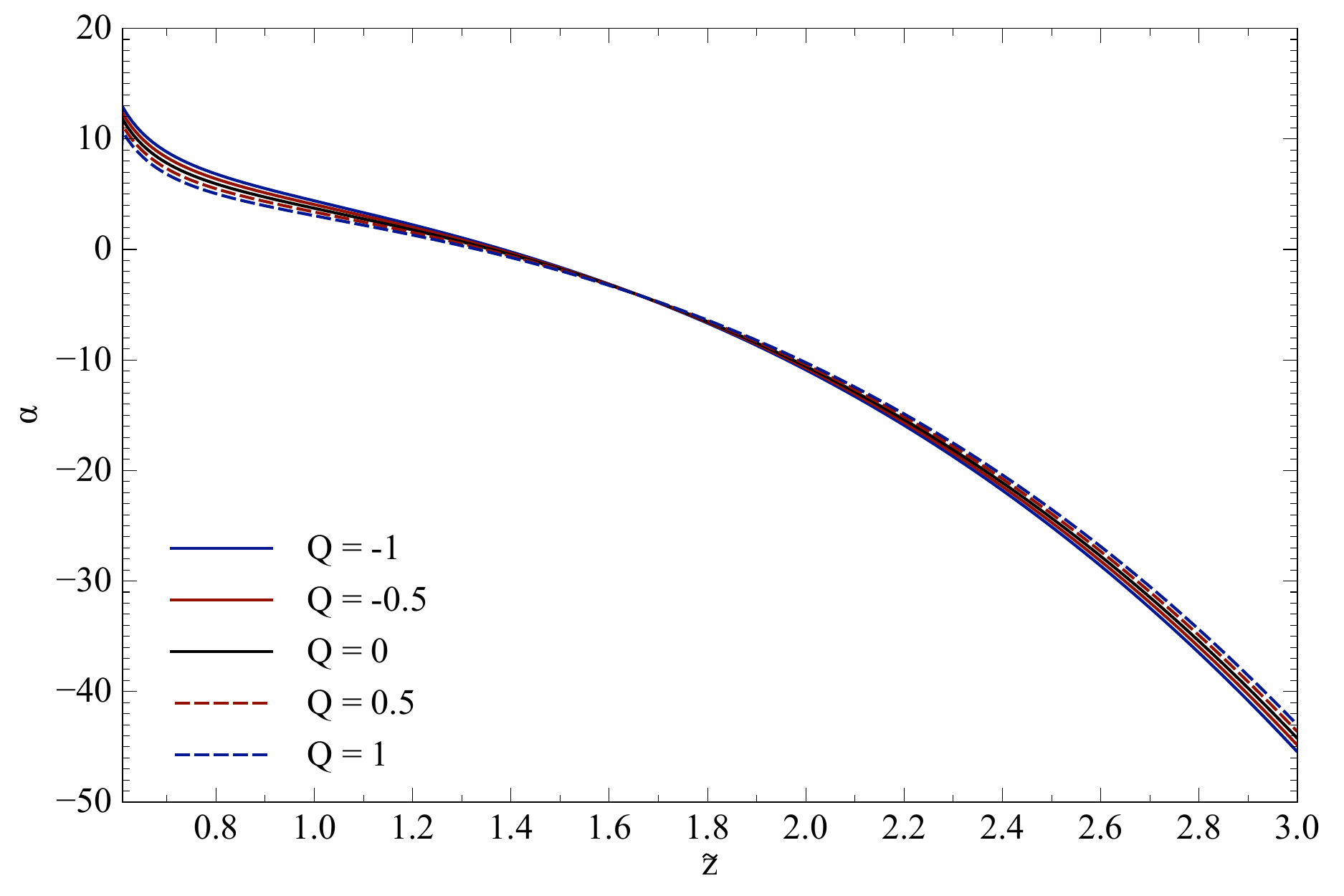}
\includegraphics[width=0.49 \linewidth]{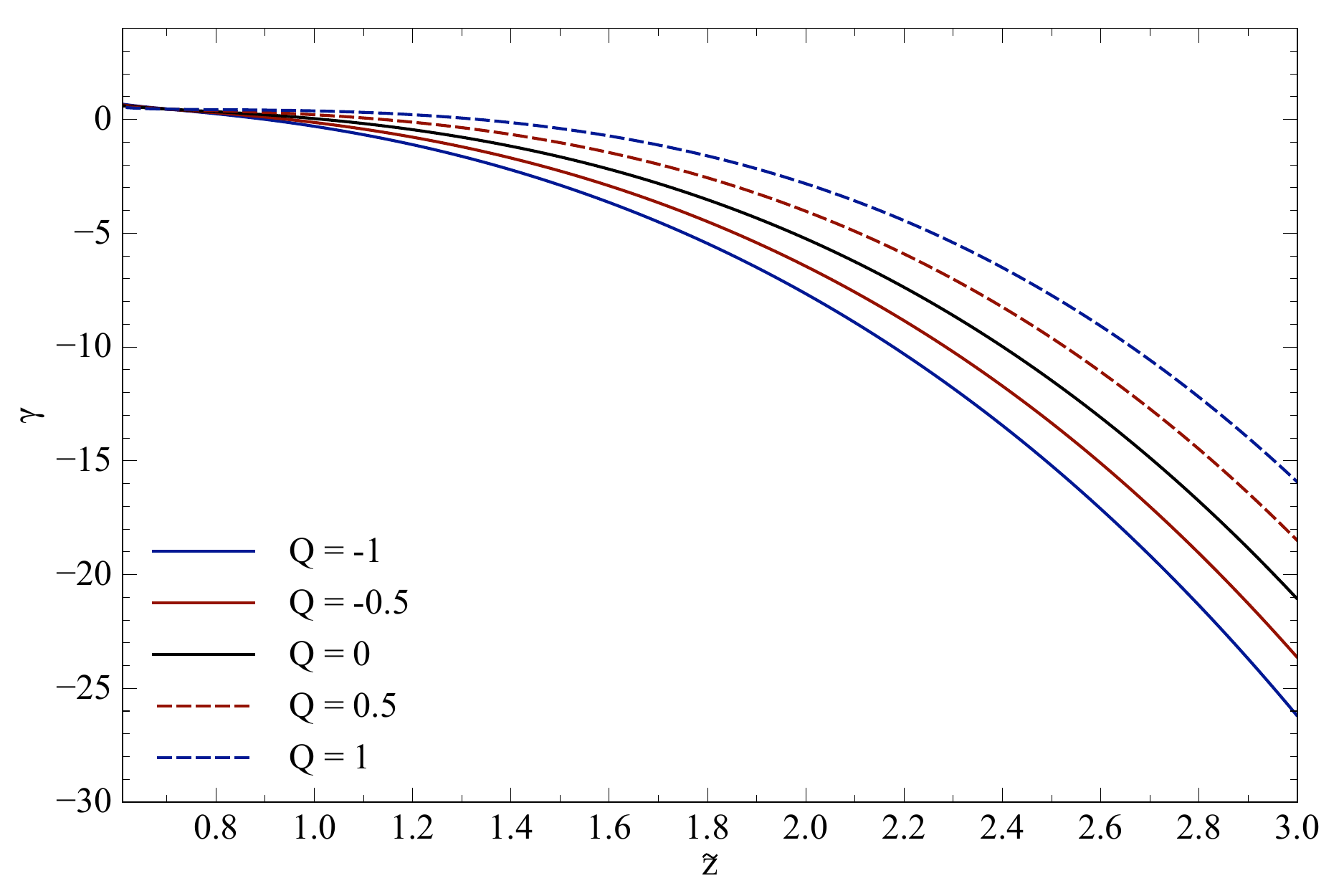}
\includegraphics[width=0.49 \linewidth]{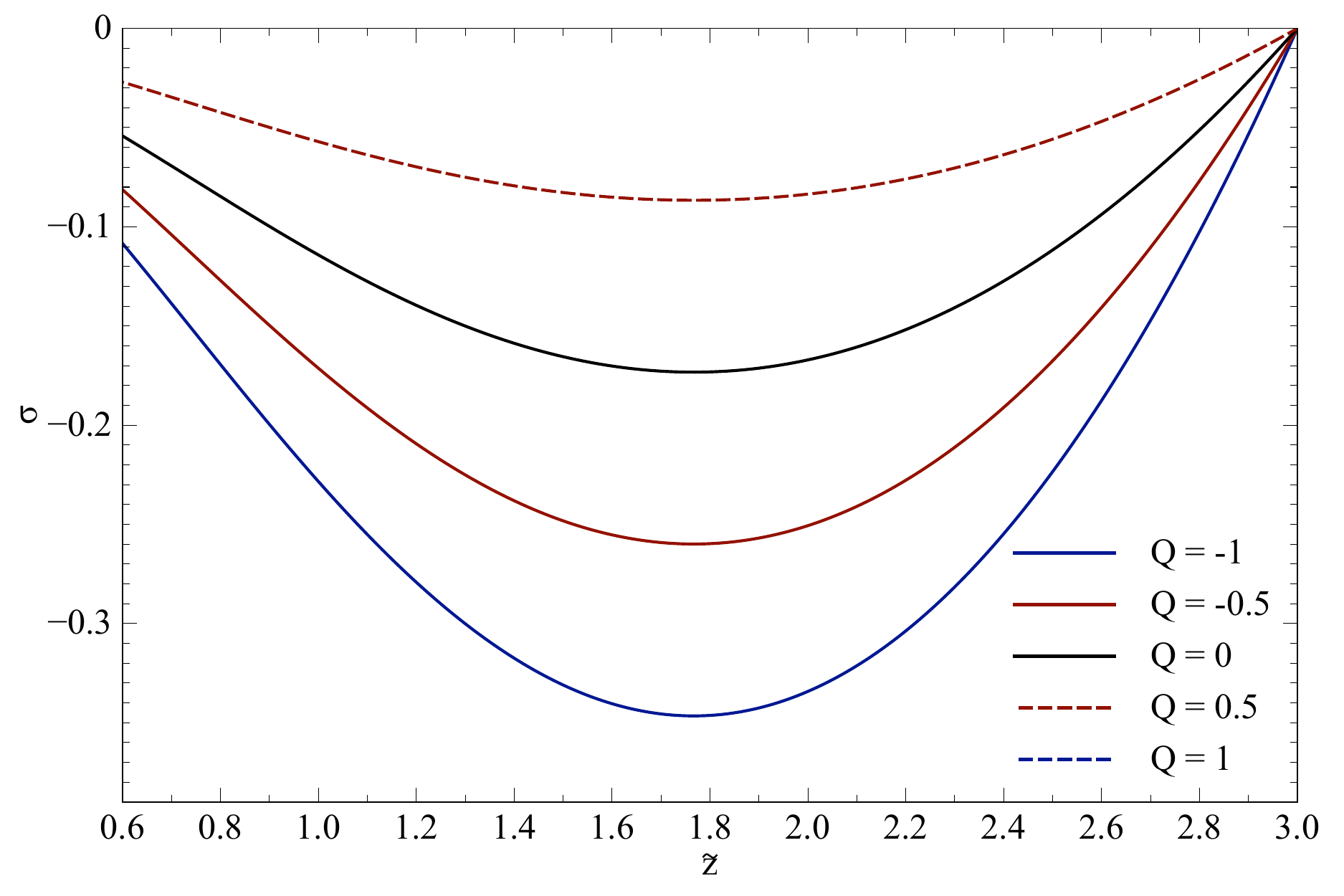}
\includegraphics[width=0.49 \linewidth]{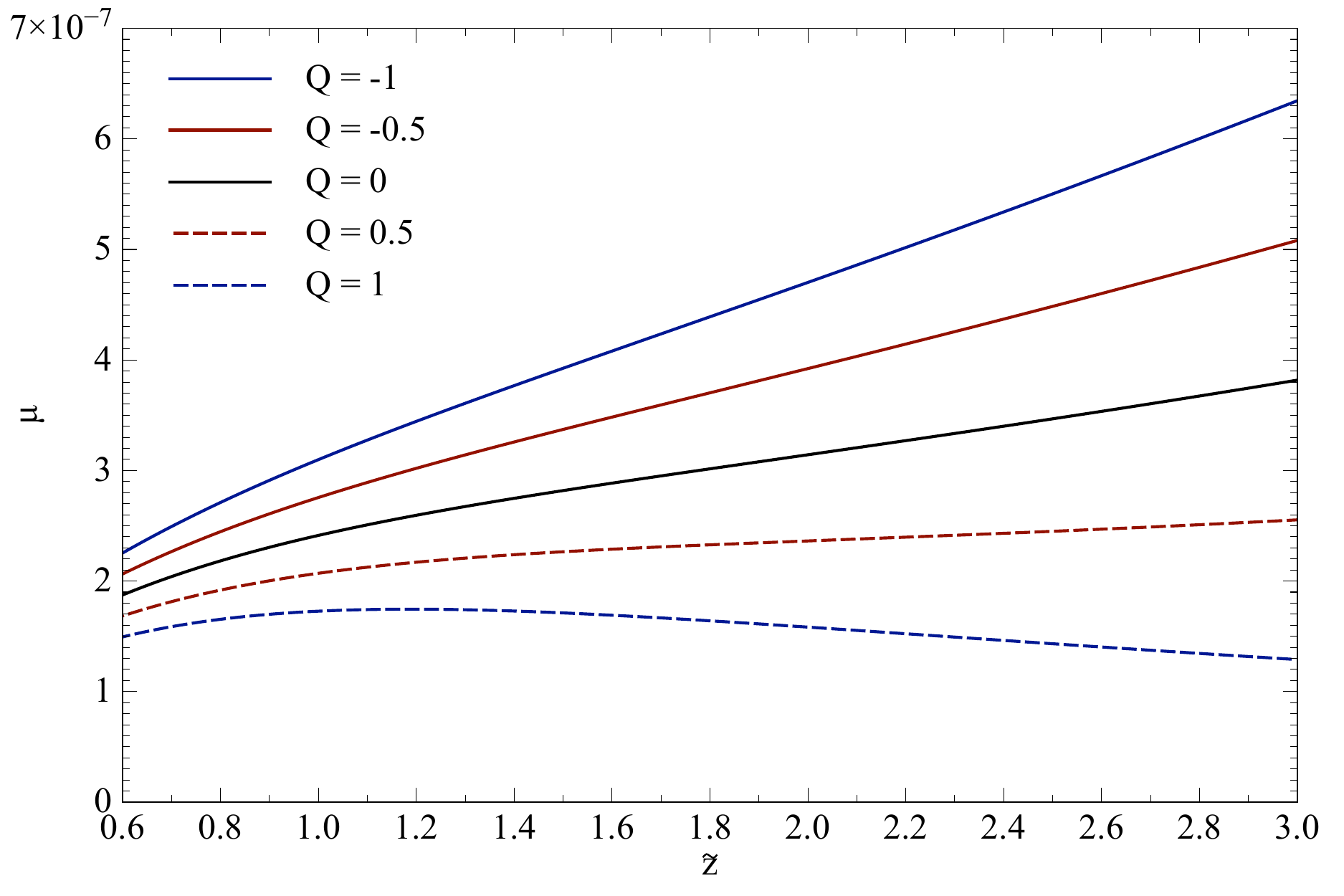}
\caption{Source functions $\alpha$, $\gamma$, $\sigma$, $\mu$ against redshift for different values of $\mathcal{Q}$. For $\mu$, $\sigma$, we take $z=3$. \label{fig:fun_Q}
}
\end{figure*}

\section{Effect of varying magnification bias }
\label{sec:q}

The magnification bias contribution to lensing in \eqref{delq}, $2{\cal Q}\kappa$, arises from the perturbation to the flux of a galaxy, relative to a galaxy at the same observed redshift in the unperturbed universe (see e.g.~\cite{Jeong:2011as, Namikawa:2011yr}). As can be seen from \eqref{eq:alpha}--\eqref{eq:mu}, the local and time-delay correlations also include the magnification bias parameter $\mathcal{Q}$, since they are affected by a magnitude-limited survey. 

The lensing contribution \eqref{delq} to the observed galaxy overdensity goes to zero when $\mathcal{Q} \rightarrow 1$ [see also~\eqref{eq:sigma}]. For galaxy surveys in H{\sc i} intensity mapping, which do not detect individual galaxies but map the total intensity of 21cm emission, ${\cal Q}=1$~\cite{Hall:2012wd}.
For optical and H{\sc i} threshold surveys, $\mathcal{Q}(z)$ will depend on specific parameters of the survey (see e.g.~\cite{Liu:2013yna}).

Here we focus on the influence of ${\cal Q}$ on the correlation function, using the reference values $\mathcal{Q}=0,\pm0.5,\pm 1$.
Figure~\ref{fig:fun_Q} shows the effect of varying $\mathcal{Q}$ on $\alpha$, $\gamma$, $\mu$, $\sigma$ ($\beta$ is independent of ${\cal Q}$), as a function of redshift. In the previous two sections, we set ${\cal Q}=0$. We can see that $\alpha$ is not affected much by the variation of the magnification bias, as expected, since it is a velocity term. The other functions all increase in magnitude as   $\mathcal{Q}$ increases. As expected, $\mathcal{Q} = 1$ gives $\sigma=0$ and the lensing term vanishes.

\begin{figure}[!htbp]
\includegraphics[width=0.49 \linewidth]{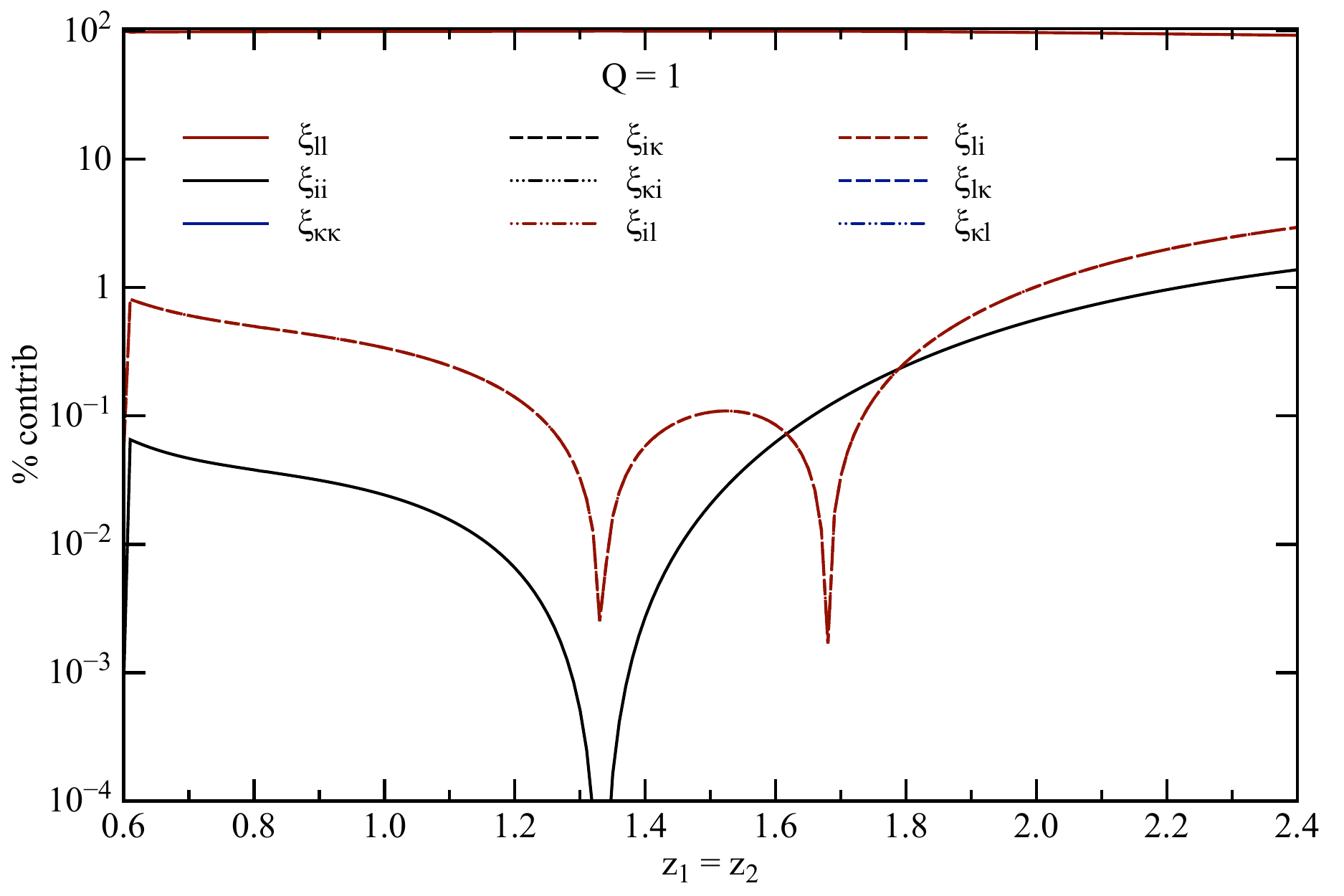}
\includegraphics[width=0.49 \linewidth]{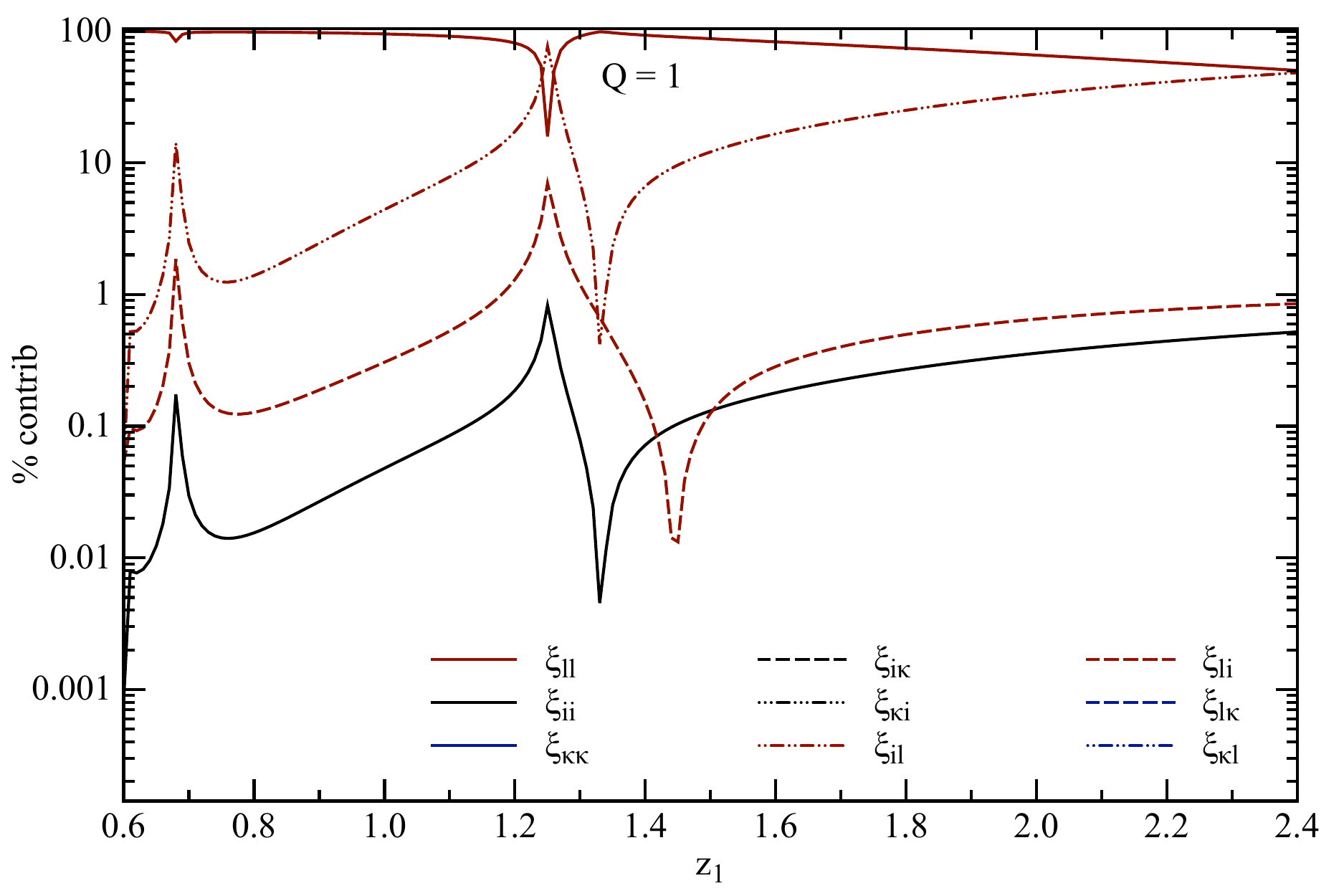}
\caption{Relative contributions of $\xi_{AB}$ with ${\cal Q}=1$. Separation semi-angle is $\theta=0.1\,$rad, with $z_1=z_2$ ({\em left}) and fixed $z_2=0.6$, varying $z_1$ ({\em right}).
\label{fig:Q1_01}
}
\end{figure}

\begin{figure}[!htbp]
\includegraphics[width=0.49 \linewidth]{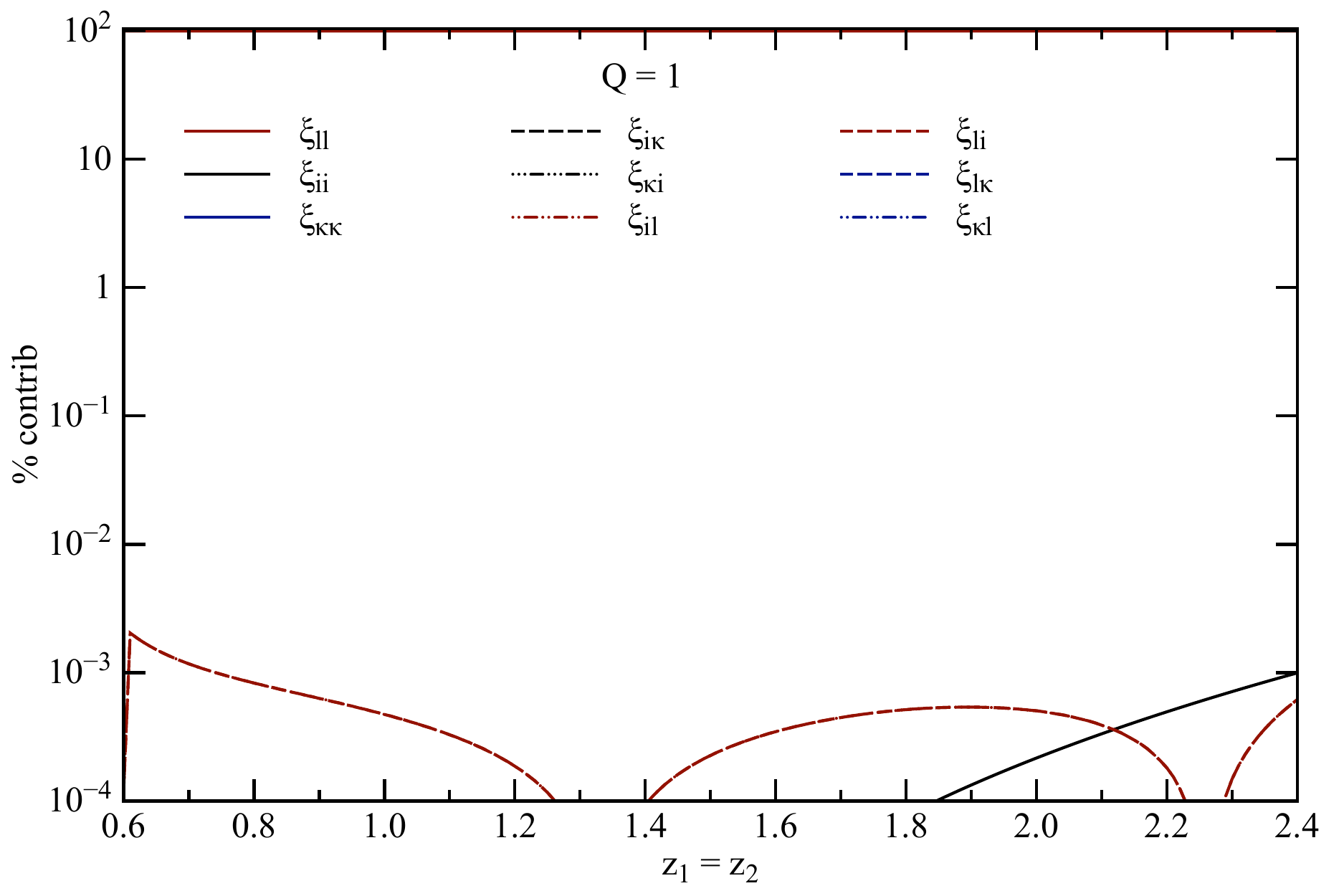}
\includegraphics[width=0.49 \linewidth]{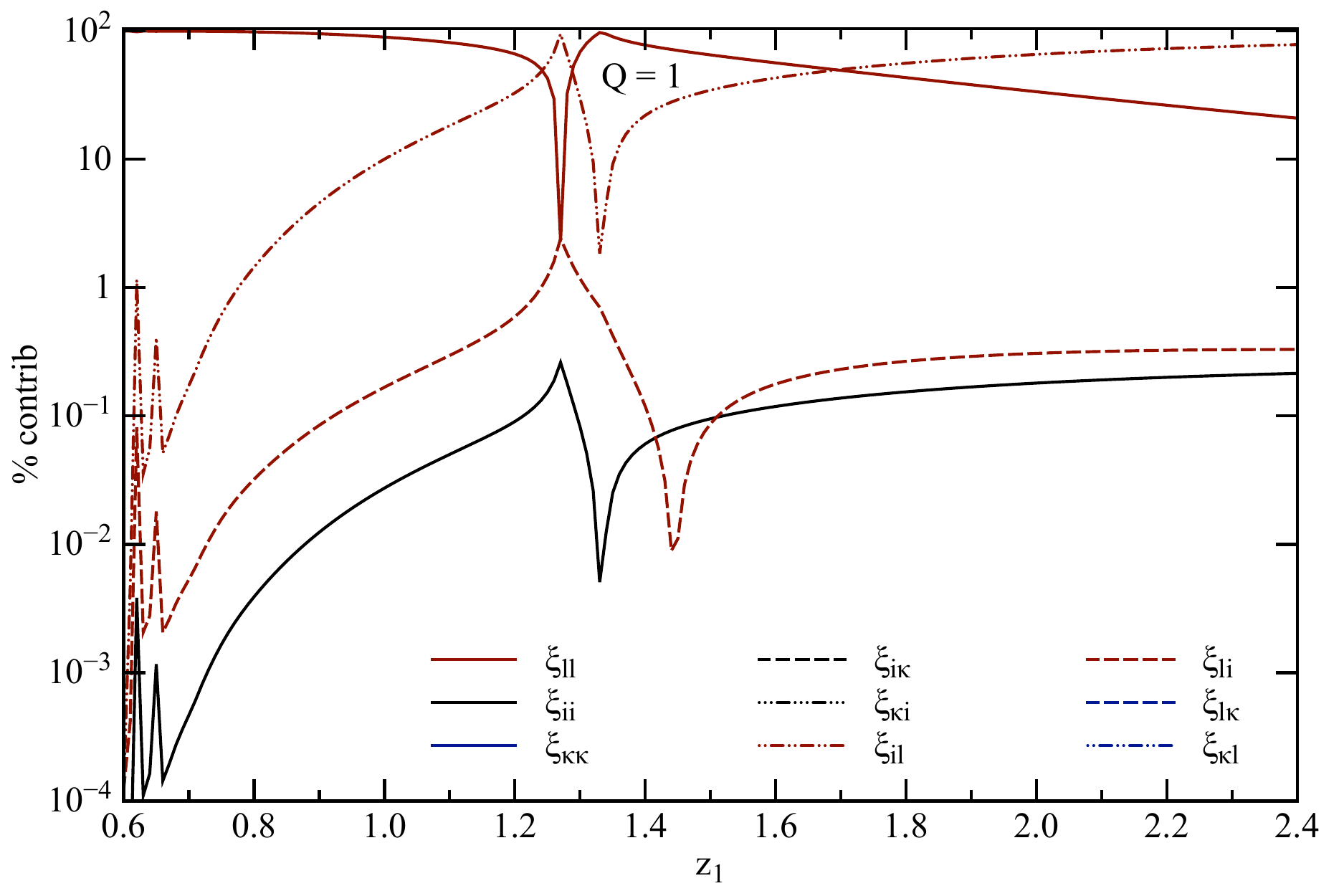}
\caption{Relative contributions of $\xi_{AB}$ with ${\cal Q}=1$. Separation semi-angle is $\theta=0.01\,$rad, with $z_1=z_2$ ({\em left}) and fixed $z_2=0.6$, varying $z_1$ ({\em right}).
\label{fig:Q1_001}
}
\end{figure}

\begin{figure}[!htbp]
\includegraphics[width=0.49 \linewidth]{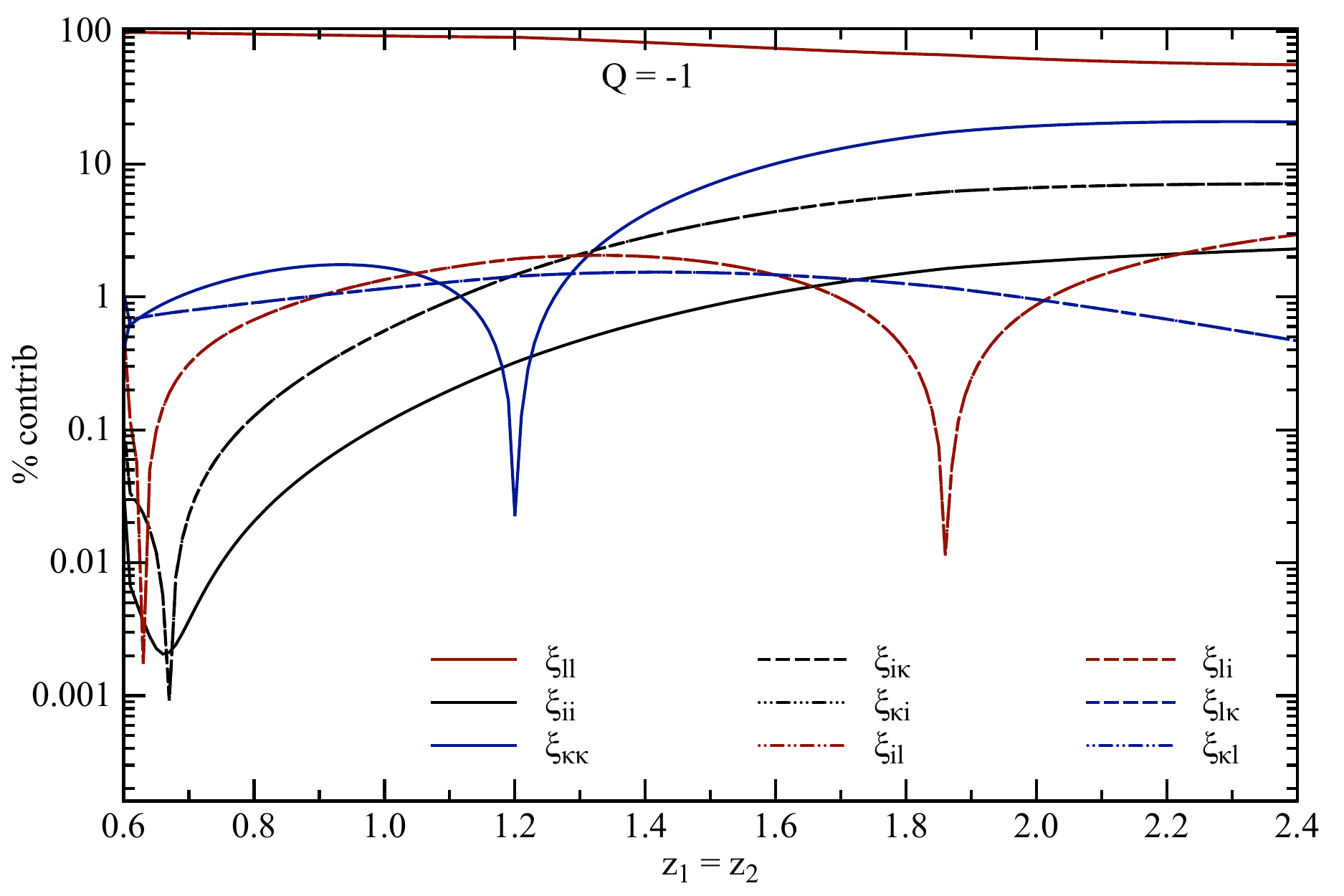}
\includegraphics[width=0.49 \linewidth]{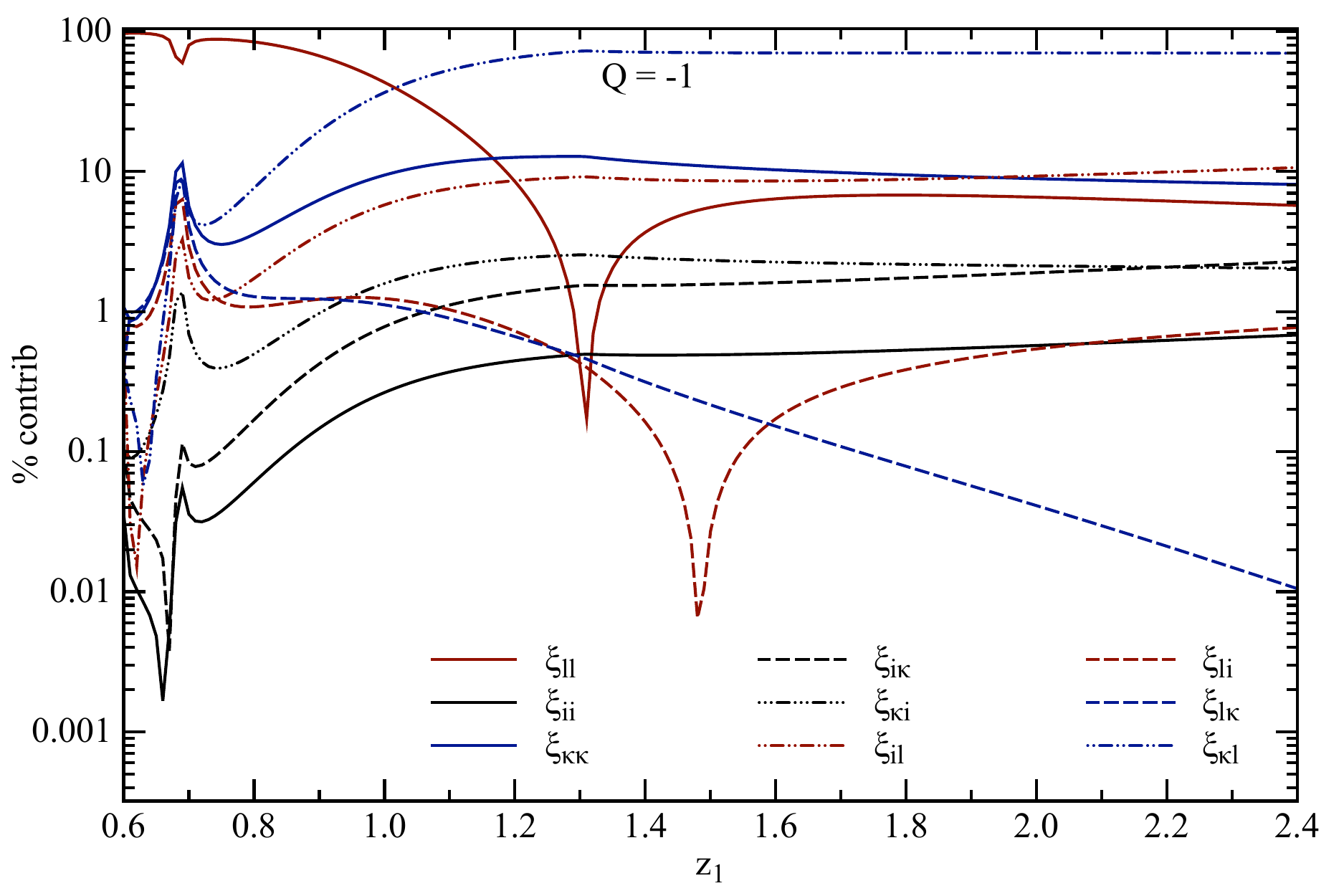}
\caption{As in Fig. \ref{fig:Q1_01}, for ${\cal Q}=-1$.
\label{fig:Q-1_01}
}
\end{figure}

\begin{figure}[!htbp]
\includegraphics[width=0.49 \linewidth]{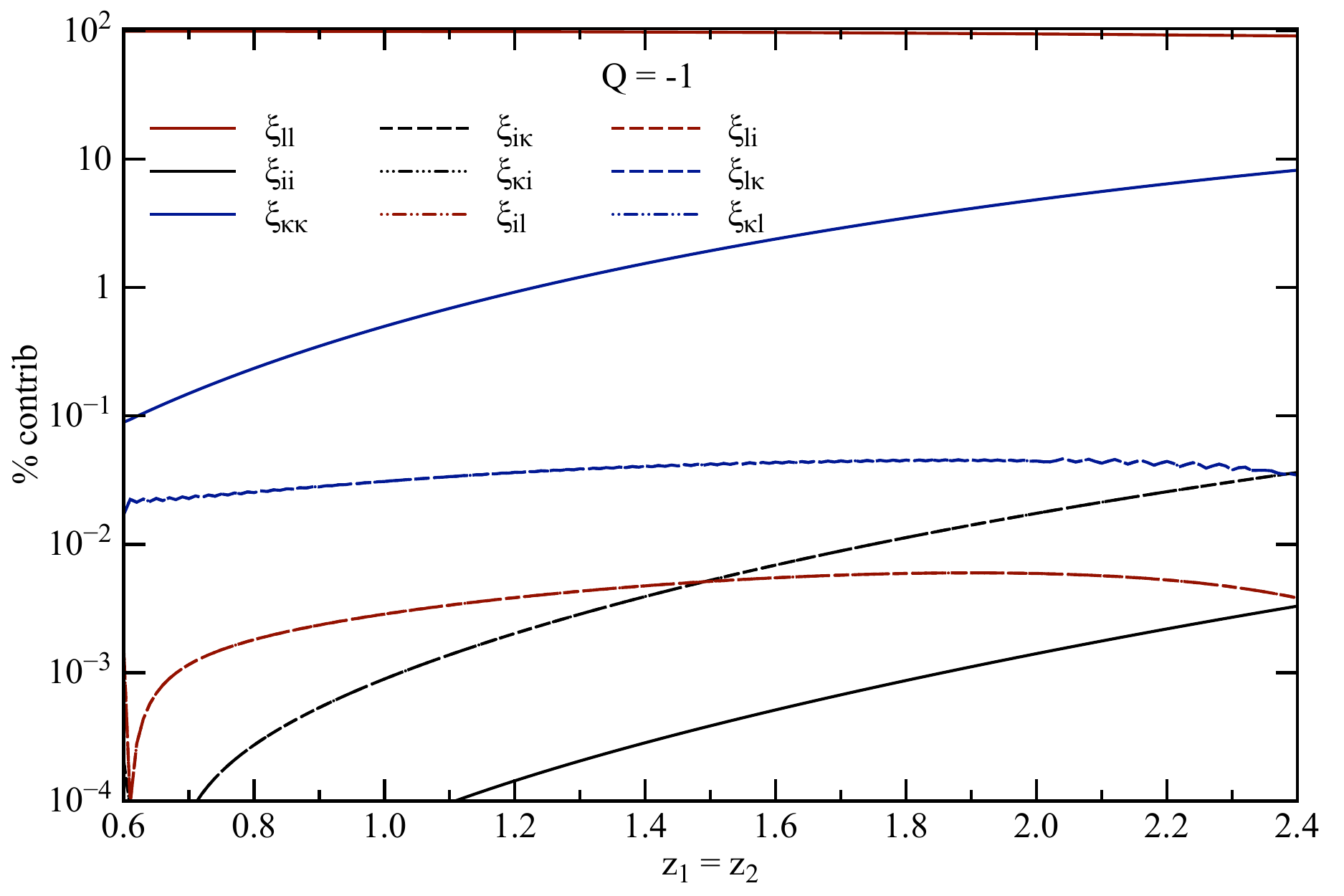}
\includegraphics[width=0.49 \linewidth]{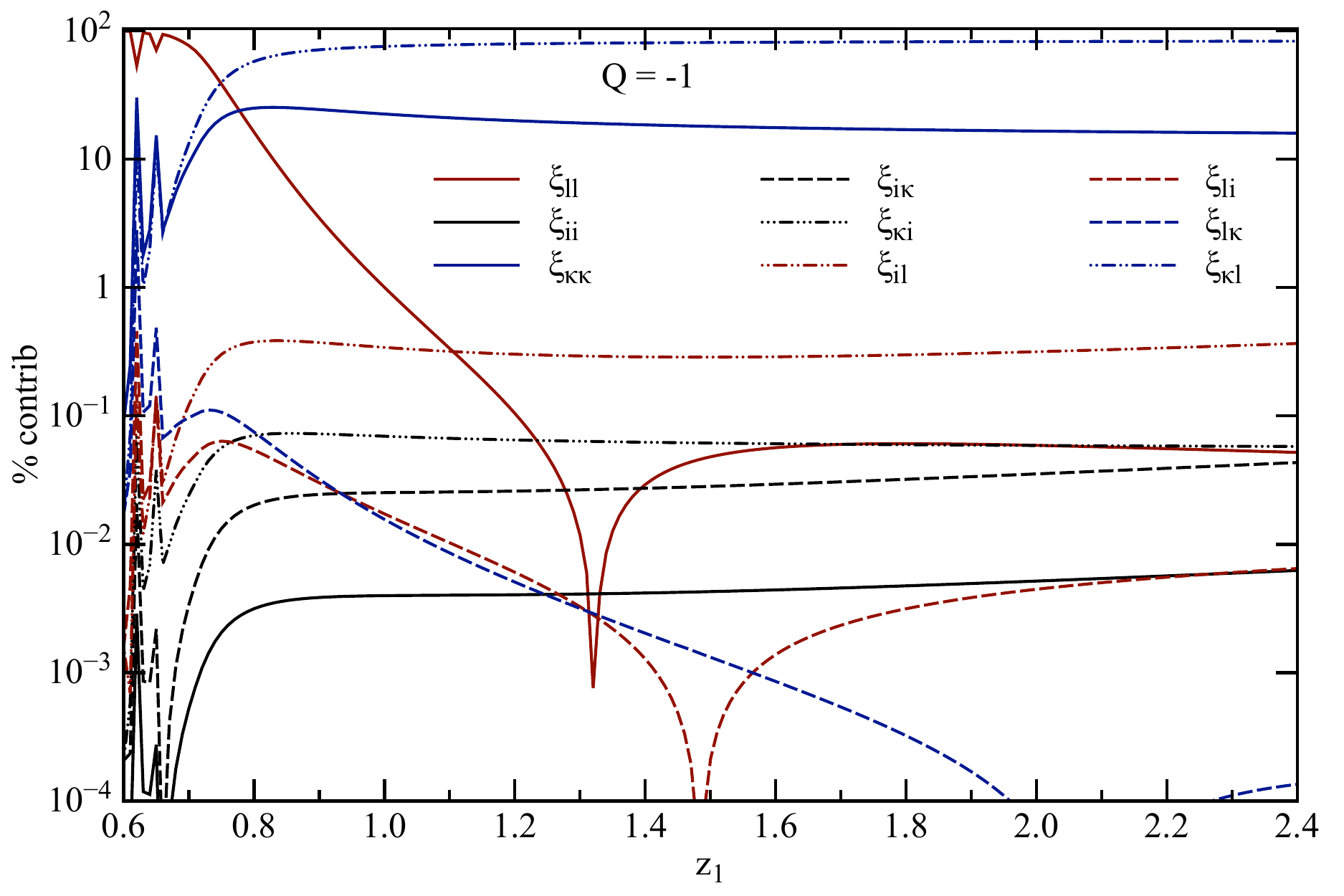}
\caption{As in Fig. \ref{fig:Q1_001}, for ${\cal Q}=-1$.
\label{fig:Q-1_001}
}
\end{figure}

In Figs.~\ref{fig:Q1_01}--\ref{fig:Q-1_001} we consider again the two cases of transverse and non-transverse separations, showing the relative contributions of the different components for two nonzero values, $\mathcal{Q} =\pm 1$. These plots can be compared with Figs.~\ref{fig:tot_contrib_perp_t01} and \ref{fig:tot_contrib_rad_01} (${\cal Q}=0$).

The results confirm that $\mathcal{Q} = 1$ kills the lensing contributions and suppresses the relativistic time-delay contributions.
In the transverse case, the local-local correlation is the only relevant component, while some integrated terms become relevant for near-radial correlations with very large linear separation, when the time-delay component becomes very important.

For $\mathcal{Q} = -1$, as expected, lensing magnification effects are larger than in the ${\cal Q}=0$ case.
In particular, in the transverse case the pure lensing contribution $\xi_{\kappa\kappa}$ reaches $\sim 20 \%$ of the total observed correlation at high redshift. 
The local-local term contributes very little in the transverse case as soon as the linear separation becomes large, and it becomes rapidly negligible in the near-radial case.
This shows that surveys with a more negative $\mathcal{Q}$ will be more powerful in detecting integrated effects and relativistic corrections. 



\section{Conclusions}
\label{sec:conclusions}
We used the fully relativistic description of the galaxy 2-point correlation function developed in~\cite{Bertacca:2012tp} to investigate the relative importance of the various contributions arising from different effects, including those not present in the standard Newtonian analysis.

First we looked at the importance of relativistic contributions to the local part of the correlation function. In the standard approach, the local correlation is given by the Kaiser approximation, possibly including Newtonian wide-angle effects. Relativistic corrections arise from the gravitational potentials (Sachs-Wolfe type terms) and from relativistic effects in the density and velocity parts of redshift-space distortions.
We considered some different configurations of the system of Figure~\ref{fig:triangle}, and we found that relativistic contributions can become non-negligible, but they contribute at most $\sim 5\%$ to the local correlation at high redshift.
More work is needed to estimate how much this will impact on precision cosmology.

Then we focused on the integrated terms arising from line-of-sight spatial and temporal perturbations, that introduce lensing, cosmic magnification and time-delay effects. These terms are typically ignored in the standard approach. However, correlations including the integrated terms can become important in most cases, but in particular in the case of large redshift separations between the two galaxies of the pair. Especially important is the lensing contribution for near-radial correlations which dominates the local-local contribution.
In the near-radial cases, the dominant term is the lensing-density one, where the redshift $z_1$ of galaxy 1 increases to high values. In some cases, the time delay-local correlation can give a significant contribution to the observed correlation. This happens because, while the local term is dominant for relatively small separations, its magnitude drops dramatically for very large separations, while the other terms do not fall off. When the first galaxy is at high-redshift, the integrals along the line of sight grow and become the dominant terms. 
Moreover, in the case of very small angular separation, the magnifying effect of the second galaxy is more evident, and also the lensing auto-correlation becomes very important, making the local part negligible.

Finally, we investigated how the magnification bias parameter $\mathcal{Q}$ affects the observed correlation. 
When $\mathcal{Q}=1$, as in the case for 21cm intensity mapping surveys, lensing effects vanish and relativistic time-delay effects almost disappear. On the other hand, surveys observing sources with a steep luminosity function will observe correlations that include larger effects from integrated terms.

Our results show that a correct modeling of large-scale galaxy clustering with wide and, more important, deep surveys, will need to take into account the effects of integrated correlations as they become dominant in the (near-)radial case. 
The importance of the integrated terms has been investigated in~\cite{Raccanelli:2015vla, Alonso:2015uua, Alonso:2015sfa, Fonseca:2015laa, Cardona:2016qxn}, showing that future single-tracer surveys can detect the lensing term at high significance~\cite{Raccanelli:2015vla, Alonso:2015uua, Cardona:2016qxn}, while multi-tracer surveys will be needed to detect the time-delay term~\cite{Alonso:2015sfa, Fonseca:2015laa}.
The correlation function for individual configurations is probably below detection, but a full analysis that averages suitably over configurations will be needed to assess the detectability of signals in future surveys.

Large-scale contributions contain additional information on the model of dark energy or modified gravity, so when investigating these scales it will be fundamental to account for all the effects to correctly model and probe the density field, and use all the constraining power available.

It is necessary to further investigate details of how these large-scale correlations depend on cosmological parameters. We leave this to a future work, together with a study on their constraining power for cosmological models.

\[\]
{\bf Acknowledgments:}\\
We  thank Guido Pettinari for useful suggestions and Nicola Bartolo, Francis-Yan Cyr-Racine, Ruth Durrer, Chris Hirata, Sabino Matarrese, Roland de Putter and Masahiro Takada  for helpful discussions.
Part of the research described in this paper was carried out at the Jet Propulsion Laboratory, California Institute of Technology, under a contract with the National Aeronautics and Space Administration.
DB and RM are supported by the South African Square Kilometre
Array Project. RM is supported by the STFC (UK) (grant no.
ST/H002774/1). RM and CC are supported by the National Research Foundation
(NRF, South Africa). DB, RM and CC were supported by a Royal
Society (UK)/ NRF (SA) exchange grant. 
Some of the numerical computations were performed on
the COSMOS supercomputer, part of the DiRAC HPC, a facility funded by STFC and BIS.

\end{document}